\documentclass[journal]{IEEEtran}
\usepackage[pdftex]{hyperref}
\usepackage{amsmath,amssymb}
\usepackage[pdftex]{graphicx}

\usepackage{tabularx} 

\newcommand{\pove}{\overline p}
\newcommand{\kove}{\overline k}
\newcommand{\cove}{\overline c}
\newcommand{\betaove}{\overline {\beta}}
\newcommand{\rbf}{\mbox{\boldmath $r$}}
\newcommand{\nablabf}{\mbox{\boldmath $\nabla$}}

\begin{document}

\title{Producing acoustic `Frozen Waves':\\ Simulated experiments with diffraction/attenuation resistant beams\\ {\em in lossy media} }

{\author{{\large{Jos\'{e}~L.~Prego-Borges, Michel~Zamboni-Rached,}} \\
{\large{Erasmo Recami, and Eduardo Tavares Costa}}}}

\maketitle

\thanks{Manuscript received ******, 2013; ****** . The authors are with the Faculty of Electrical Engineering of the State
University of Campinas at Campinas (SP), Brazil (e-mail: mzamboni@dmo.fee.unicamp.br); and Erasmo Recami is also with
INFN-Sezione di Milano, Milan, Italy, and the Faculty of Engineering of the Bergamo state University, Bergamo, Italy.
Digital Object Identifier XXXXXXXXXXXXXXXXXXX. This work was partially supported by CNPq, Brazil, under grant no. 500364/2013-3 (JLPB), and by FAPESP, Brazil, under grant no. 2013/12025-8 (ER), besides by INFN, Italy. Their support is acknowledged.}
\markboth{ }{Prego-Borges \MakeLowercase{{\rm et al.}}: Acoustic `Frozen Waves' in lossy media}


\begin{abstract}
The so-called Localized Waves (LW), and the ``Frozen Waves" (FW), have arisen
significant attention in the areas of Optics and Ultrasound, because of their surprising energy localization
properties. \ The LWs resist the effects of diffraction for large distances, and possess an interesting self-reconstruction
({\em self-healing\/}) property (after obstacles with size smaller than {\em the antenna's\/}); while the FWs, a
sub-class of theirs, offer the possibility of arbitrarily modeling the field longitudinal intensity pattern inside a
prefixed interval, for instance $0 \leq z \leq L$, of the wave propagation axis. \ More specifically, the FWs are
localized fields ``at rest'', that is, with a {\em static} envelope (within which only the carrier wave propagates), and
can be endowed moreover with a high transverse localization.

In this paper we investigate, by simulated experiments, various cases of generation of ultrasonic FW fields, with the frequency of
$f_0=1\,\text{MHz}$ in a water-like medium, taking account of the effects of {\em attenuation}. We present results of FWs for distances up to
$L=80\;$mm, in attenuating media with absorption coefficients $\alpha$ in the range $70\le \alpha \le 170\;$dB/m.

Such simulated FW fields are constructed by using a procedure developed by us, via appropriate finite superpositions
of monochromatic ultrasonic Bessel beams.

We pay due attention to the selection of the FW parameters, constrained by the rather tight restrictions imposed by
experimental Acoustics, as well as to some practical implications of the transducer design.

The energy localization properties of the Frozen Waves can find application even in many medical
apparatus, such as bistouries or acoustic tweezers, as well as for treatment of diseased tissues (in
particular, for the destruction of tumor cells, without affecting the surrounding tissues; besides for kidney stone
shuttering, etcetera).

\end{abstract}


\section{Introduction}
\hyphenation{non-diffracting}

The phenomena of diffraction, dispersion and attenuation are physical effects that usually perturb the propagation of
waves. Diffraction produces a gradual spatial spreading of the wave;
while dispersion produces a temporal spreading.  Dispersion, for example, is present whenever the refraction index of the
medium depends on the frequency, as in Optics; this produces a progressive increase in the temporal width of the propagating
pulses.

Attenuation reduces the amplitude of a propagating wave, by transforming, roughly speaking, part of its energy into heat.

To circumvent these difficulties, which affect the development of any technological applications, increasing interest was paid to
the Non-Diffracting Solutions to the wave equations, also known as ``Non-Diffracting Waves" or {\em Localized Waves} (LW). These waves were first theoretically predicted [see, e.g., Ref.\cite{CH}], and then ---to confine ourselves to the simplest case of the Bessel beams--- experimentally produced in a series of papers, among which Refs.\cite{sheppard1,sheppard2,durnin1987}.

Indeed, the LWs are solutions to the wave equations capable of resisting the effects of diffraction, and in some cases even of attenuation, at least up to a certain distance ({\em depth of field\/}). \  Such fields, being solutions to the wave
equations, can find application in diverse technological areas, in Optics\cite{hugo2008,hugo2013,recami2009,JSTQE,tarcio2012}, in
Acoustics\cite{lu1992,lu1992b,lu1995,pregoI,hugo2008,hugo2013}, in Geophysics\cite{Sanya}, and so on; besides playing interesting theoretical roles,
even in special relativity\cite{ReplyToSeshadri,chapterII,saari2004,zamboni2008}, quantum mechanics\cite{schX}, etc. \ In a very large number of theoretical
and experimental works, such soliton-like solutions to the {\em linear} wave equations have been shown to be endowed with
peak-velocities $V$ ranging from $0$ to $\infty$; even if they have been extensively studied\cite{hugo2008,hugo2013,recami2009} mainly
for their peculiar properties, like their self-recontruction (``self-healing") after obstacles with size much larger than the wavelenght, provided that
it be smaller than their antenna, and not at all for their speed.

The LWs can be either Localized Beams or Localized Pulses. For reviews, one may consult the already quoted
Refs.\cite{hugo2008,recami2009,JSTQE,hugo2013} and refs. therein.  Rather famous became the ``superluminal" pulses,
called X-{\em shaped waves}\cite{lu1992,lu1992b,PhysicaA,saari1,saari2}.

If we refer more specifically to the application of LWs in the areas of Acoustics, one has first to recall the works by
J.-y.Lu et al.\cite{lu1992,lu1992b,lu1995,lu1994,lu1994b,lu1997,lu2000}, whose principal interest became soon that
of applying the ultrasonic X-shaped pulses (endowed with a {\em supersonic} peak-velocity, in this case) for medical imaging. Other
research related with ultrasonic localized pulses can be found ---besides in pioneering papers
like\cite{McLeod1,McLeod2,UltrasAxic}---
also in\cite{fox2003,castellanos2010,castellanos2010b}. In connection with single Bessel beams, one can recall for instance
papers like Refs.\cite{hsu1989,holm1998,nowack2012}; while, with regard to the use of annular transducers, one can recall for
example Refs.\cite{lu1990,eiras2003,aulet2006,moreno2010,calas2010,castellanos2011}. Among the many existing papers, let us
mention Refs.\cite{domell1982,foster1989,fox2002a,fox2002b,martinez2000} related with annular {\em arrays}; or
Refs.\cite{martinez2000,martinez2001,akhnak2002,ullate2002,godoy2006} for {\em segmented} annular arrays; or
Refs.\cite{marston2007,mitri2010,mitri2011,marston2011} for Bessel beam superpositions suitable in the case of ultrasound wave scattering by spherical objects.

As said before, LWs can have peak-velocities $V$ ranging from $0$ to $\infty$. \ There exist, therefore, also the
{\em subluminal} ones (properly speaking {\em subsonic}, in the case of Acoustics), which have been investigated mainly in
Ref.\cite{zamboni2008}, after some pioneering or preliminary works.\cite{lu1995b,sheppard2002,besieris2008}

In such a ``subsonic" area, some publications\cite{zamboni2004,zamboni2005,zamboni2006,zamboni2011}, most of them
in Optics, had recourse to superpositions of monochromatic Bessel beams with the aim of modeling, in an arbitrary
way, the shape of the LWs. Indeed, the (subluminal) localized pulses reduce to (monochromatic) beams when their peak-velocity
tends to zero ($V \rightarrow 0$). It is just in the case of such waves ``at rest" ---that is, with a {\em static} envelope,
within which only the carrier wave propagates--- that it resulted to be possible to model the LW shape in a pre-fixed way
(and inside a chosen space interval). Such peculiar localized waves were called {\em Frozen Waves} (FW) by us: see
Refs.\cite{zamboni2004,zamboni2005,zamboni2006,zamboni2010}, and Ref.\cite{chapterII}, just appeared as Chap.1 in the new
book\cite{hugo2013}. \ The FWs have been experimentally generated for the first time, in 2012, in Optics\cite{tarcio2012};
while computer simulations of the experimental production of ultrasonic FWs have recently yielded promising
results\cite{pregoI}.

We initially developed the theory of FWs for lossy-less media\cite{zamboni2004}, but subsequently was extended by us for absorbing
media\cite{zamboni2006}: Where it showed to be able to spatially model non-diffracting beams in order to obtain {\em beams resistant to both diffraction and attenuation}. This was performed by following the procedures exploited in Ref.\cite{zamboni2006}, which allows individuating the appropriate finite superpositions of monochromatic zero-order ($\mu=0$) Bessel beams, with different longitudinal wavenumbers $\beta_m$.

It is possible, of course, to conceive superpositions of Bessel beams of higher order ($\mu>1$), and then construct
localized ``tubes'' of energy along the $z$ axis, as in Refs.\cite{zamboni2010,zamboni2011}. However, this endeavor
---a priori, quite possible in Optics, by utilizing the versatility of the spatial light-modulators--- becomes a
technological challenge in the ultrasound case: Since it would probably require a further subdivision of the transducer
rings (that we adopted in our previous paper\cite{pregoI}, and will be considered below)
into small arc segments, in order to get what is called a segmented annular
array\cite{martinez2000,martinez2001,akhnak2002,ullate2002,godoy2006}. In fact, a radiator structure like that seems to be
needed because of the azimuthal dependence of the fields along the $z$
axis,\footnote{In this paper we use cylindrical coordinates, $\rho,\phi,z$.} through the
phase $e^{i\mu\phi}$. \ This is not a simple problem, if one bears in mind that it would involve a second \textit{angular}
sampling process, besides the main sampling process, SP, required to generate the FW patterns: See Sec.6 below. \ [And a
wrong sampled structure of a segmented annular array could completely distort the entire FW to be generated (an effect
expected to occur even in the simple case of $\mu=0$ Bessel beams)].  \ To our knowledge, the study of the SP effect for FWs is still largely unexplored. \ Moreover, a chosen segmented annular transducer has to usable, in principle, for the generation of several FW patterns: That is to say,
one and the same particular segmented annular structure of the radiator ought to be appropriate for all the FWs to be created.

We feel, as a consequence, that questions like the use of higher order Bessel beams must wait for a deeper understanding
and control of the simple $\mu=0$ case; and in this work we shall confine ourselves to zero-order Bessel beams.

The main practical issue is investigating the suitable ultrasonic transducers: This is of course a key point in order
to make our acoustic FWs realizable. \ Another related issue is the `distorting' effects possibly introduced by the responses of
the individual radiator annuli: Namely, the effects of signal attenuation and delay produced by the yet unknown electrical/mechanical transfer function of each ring of the (piezoelectric) transducer. Once again, we do
not addressed this point here, since this work is chiefly devoted to the theoretical/simulation aspects of FWs. \ However we shall comment on a possible solution to this problem at the end of the paper. \ At last, also the number of electronic channels involved in the production of the FWs is an important topic during the design process; but, with today's availability of cheap electronics and ready-to-use multichannel electronic front ends, this problem looks much less severe.\\

\hyphenation{de-ve-lop-ment}
Purpose of this paper is contributing to the creation of {\em Frozen Waves} in the ultrasound sector, by presenting a series of computer simulated experiments for FWs in a water-like medium. We shall show a modeling of the non-diffracting beams to be possible, so as to overcome the effects of diffraction and attenuation all along their depth of field. {\em Inclusion of the effect of attenuation} is rather important for instance when wishing to construct acoustic FWs inside the human body, having in mind the most useful applications of theirs\cite{patent} that one can imagine at the moment.

To this aim, Sec.2 introduces some generalities on attenuation of ultrasound in fluids, in particular in water. The modeling of
FWs in an absorbing medium is presented in Sec.3 (for a review of the general methodology for the generation of FWs in
non-attenuating media, cf., e.g., Refs.\cite{pregoI,zamboni2005,hugo2008,hugo2013}). Section 4 discusses the acoustic restrictions
on the values for the parameters to be used for ultrasonic FWs: As we shall see, the main limitation in the acoustic regime is related to the low values of the ratio frequency/speed of sound in the medium (a problem that is not that severe in Optics).

A brief introduction to the impulse response (IR) method, adopted by us for the simulation of acoustic FWs, is made in Sec.6. We shall describe therein the inclusion of the medium attenuation effect via a linear model of the absorption that affects the initial,
non-attenuated IR signal at each particular point of space.

In Sec.7 we show the results of our simulated experiments for three different ultrasonic FWs, in a water-like medium (and
including, as we said, the effect of absorption). In these ``experiments", all performed with the frequency
$f_0=1\,\text{MHz}$, we actually contemplated three absorption scenarios, with various absorption coefficients $\alpha$ in the
range $70\le \alpha \le 170\,\text{dB/m}$.

This paper ends with conclusions, and some new ideas for future developments.

In an Appendix we further discuss how a proper choice of parameters like $L$ can help in the experimental generation of the FWs.

\section{Attenuation of Ultrasound in Water: Generalities}

The phenomenon of attenuation of ultrasound in fluids may in general be associated to three type of losses or relaxation
mechanisms: i) heat conduction, ii) viscosity, and iii) internal molecular processes.
In polar liquids such as water, the thermal relaxation loss does not account for the excess of absorption observed in reality:
which is in fact three times bigger than the classical absorption value. Then, the main causes of the attenuation in normal water (not sea water) are due to the other causes, such as the viscosity and internal molecular losses.

Viscous losses happen whenever there is a relative motion between adjacent portions of the medium. This is the case for example when a wave of sound (a longitudinal wave) produces compressions and expansions in the medium while propagating. \ This phenomenon causes a kind of diffusion of the momentum of the wave, due the molecular collisions between adjacent regions with different velocities. The whole processes can be measured by the shear-viscosity coefficient $\eta$.

On the other hand, the losses produced by the internal molecular processes may be attributed to a structural change in the fluid volume, which occurs at a microscopic level. The theory that explains this mechanism is the 1948 Hall's theory of structural
relaxation\cite{hall}. That theory basically claims the water to have two energy states: One with a lower energy (the normal state), and one, with a higher energy, in which the molecules have a more packed structure. Under normal conditions, most of the molecules are in the first state of energy. However, the passage of a compressional wave produces the transition of the molecular status to
a more strictly packed state. \ Such a process (and its reversal) lead to a relaxation mechanism and to dissipation of the wave energy. This can be accounted for by a non-vanishing bulk viscosity coefficient,  $\eta_{\beta}$.

The above-mentioned causes of attenuation in normal water (which is the case assumed in this paper) can be represented by an
overall relaxation-time constant $\tau_s$, on the basis of a linear analysis\cite{kinsler2000} of the Navier-Stokes equation.  From it the following (lossy) Helmholtz equation for the acoustic pressure can be derived:
\begin{equation}
\label{eqn_1}
  {\nablabf^2 \pove + \kove^2 \pove} = 0 \; ,
\end{equation}

quantity
\begin{equation}
\label{eqn_2}
  \kove = k + i\alpha_s = \frac{\omega}{c\,\sqrt{1 - i\,\omega \tau_s}}\ ,
\end{equation}

\noindent being the complex wave vector\footnote{In this paper we use overlined letters to denote complex variables.}.

Equations \ref{eqn_1} and \ref{eqn_2} are only valid when the fluid is assumed to be continuous, what can be summed up by imposing the condition $\omega\tau_s \ll 1$. A more useful approximation for $\alpha_s$, derived from Eq.\ref{eqn_2}, is
\begin{equation}
\label{eqn_3}
  \alpha_s \approx \alpha = \frac{\omega^2\,\tau_s}{2c} = \frac{\omega^2}{2\rho_0 c^3} \left( \frac{4}{3}\eta + \eta_{\beta} \right)
\end{equation}

Here, the viscosity coefficients $\eta$ (shear viscosity) and $\eta_{\beta}$ (bulk viscosity) are both in Pa$\times$s. On the other hand, the complex wavenumber $\kove$ and the complex speed of sound $\cove = c_R+ic_I$ are related by the expression
\begin{equation}
\label{eqn_4}
  \kove = k + i\alpha = \frac{w}{\cove} \ ; \ \ \text{with} \ \ k \approx \frac{2\pi}{\lambda}
\end{equation}

Using these quantities and assuming a monochromatic excitation $\omega_0$, a damped plane-wave solution of equation (\ref{eqn_1}) propagating along the $z$ axis can be expressed as:
\begin{align}
\label{eqn_5}
  \pove &= P_0\, e^{i(\kove z-\omega_0 t)}\\
             &= P_0\, e^{-\alpha z} e^{i(kz-\omega_0 t)}\nonumber \; ,
\end{align}

\noindent where the coefficient $\alpha$, measured in Np/m, finally accounts for all the effects of attenuation of the fluid on
the traveling wave.

\section{Diffraction/attenuation resistant beams: Frozen Waves (FW) in Absorbing Media}

As we discussed in the previous Section, the effects of attenuation of ultrasound in normal water can be represented by
a decaying exponential term $e^{-\alpha z}$ in the solution, Eq.(\ref{eqn_5}), of the lossy wave equation (\ref{eqn_1}).

In order to model our acoustic FWs in an attenuating medium, we have to modify the language previously used by us for {\em optical} FWs propagating in lossy media, by taking in mind that, given an
excitation frequency $\omega_0$, in an acoustic absorbing medium a plane wave possesses a complex wavenumber $\kove$ obeying relationship (\ref{eqn_4}) and has the complex sound speed $\cove$.  We are therefore going to summarize the method in Refs.\cite{zamboni2006,zamboni2011}.

An appropriate continuous superposition of such plane waves\cite{hugo2008,zamboni2005}, with wave vectors laying on the surface of a cone with angular aperture $2\theta$, will form a Bessel beam\cite{hugo2008,hugo2013}:
\begin{equation}
\label{eqn_6}
  \boldsymbol{\psi} = J_0(\kove_{\rho}\rho)\; e^{i(\betaove z-\omega_0t)} \ ,
\end{equation}

\noindent which obeys the dispersion relation
\begin{equation}
\label{eqn_7}
  \kove_{\rho}^2 + \betaove^2 = \frac{\omega_0^2}{\cove^2}
\end{equation}

Here the transverse, $\kove_{\rho}=\kove \sin(\theta)$, and longitudinal, $\betaove =\kove \cos(\theta)$, spatial components of the complex wavenumber
$\kove = k+i \alpha$ satisfy the following relations:
\begin{align}
\label{eqn_8}
  \kove_{\rho} &= k\sin(\theta) +i\, \alpha\sin(\theta) = k_{\rho R}+i k_{\rho I}
  \ \\
  \betaove &= k\cos(\theta) +i\, \alpha\cos(\theta) = \beta_R+i\beta_I \nonumber
\end{align}

\noindent From them it is possible to derive that
\begin{equation}
\label{eqn_9}
  \frac{\beta_{R}}{\beta_{I}} = \frac{k_{\rho R}}{k_{\rho I}} = \frac{k}{\alpha}
\end{equation}

We can then write the expression of the Bessel beam in terms of the complex components of $\kove_{\rho}$ and $\betaove$ as
\begin{equation}
\label{eqn_10}
  \boldsymbol{\psi} = J_0[(k_{\rho R}+i k_{\rho I})\rho]\; e^{i(\beta_R+i\beta_I)z}\; e^{-i\omega_0t}
\end{equation}

\begin{figure}[t]  
\centering
\includegraphics[width=2.25in]{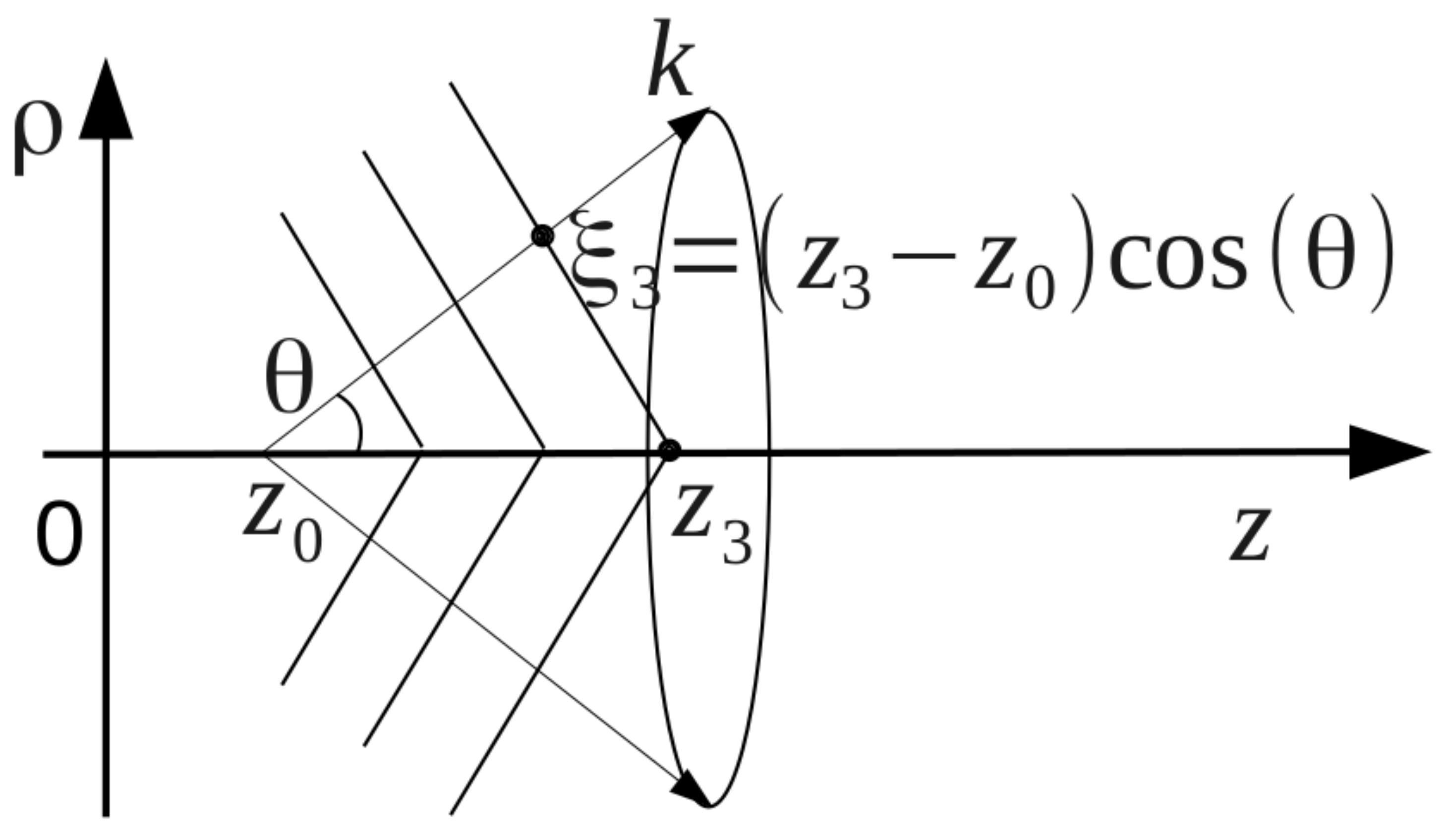}
\caption{Illustration of the interference of the plane waves forming a Bessel beam.}
\label{fig_1}
\end{figure}

It is also interesting to note that, although a plane wave has the attenuating term $e^{-\alpha z}$ [cf. Eq.(\ref{eqn_5})], a  Bessel beam (as well as a superposition of them) has the attenuating factor $e^{-\alpha \cos(\theta) z}$, which in fact is smaller than the former. This behavior can be seen by observing Figure \ref{fig_1}, which illustrates the interference of the plane waves belonging to a Bessel beam (and whose wave vectors, as is known, stay on the surface of cone with axicon angle $\theta$). Notice how, in the same amount of time, the Bessel has advanced a distance  $z_3-z_0$, while the corresponding plane-wave front has only traveled a distance $\xi_3=(z_3-z_0)\,cos(\theta)$, measured from the vertex of the cone. \ That is, the Bessel beam has traveled a longer distance than the single plane wave, or, in other words, along the same distance the attenuation of the plane wave is bigger by the factor $1/\cos(\theta)$ than the Bessel beam's.

Let us now create a FW, endowed with the wished longitudinal intensity pattern $|F(z)|^2$, in the lossy medium, around the propagation axis $z$ (that is, on \;$\rho=0$). \ The envelope function $F(z)$ can be arbitrarily selected inside the prefixed interval $0\leq z \leq L$, provided that the diffraction limits are respected. \
To such a purpose, let us construct the following finite superposition
of co-propagating $2N+1$ Bessel beams:
\begin{equation}
\label{eqn_11}
  \boldsymbol{\Psi}(\rho,z,t) = e^{-i\omega_0 t} \sum\limits_{m=-N}^{N} A_m
  J_0{k_{\rho m} \rho} \; e^{i\beta_m z}
\end{equation}

where

$$\kove_{\rho m} \equiv k_{\rho R_m} + ik_{\rho I_m}, \ {\rm and } \ \betaove_m \equiv \beta_{R_m} + i \beta_{I_m}$$

obey, for each value of $m$, the equations (\ref{eqn_7}) and (\ref{eqn_9}).

Lets us write the real component of $\betaove_m$ in the form

\begin{equation} \label{eqn_12}
  \beta_{R_m} = Q + {\frac{2\pi m}{L}}
\end{equation}

and recall that, once $\beta_{R_m}$ has been chosen, quantities $\beta_{I_m}$ and $\kove_m$ are automatically defined through
Eqs.(\ref{eqn_9}) and (\ref{eqn_7}), respectively.  In Eq.(\ref{eqn_12}) the parameter $Q$ has an arbitrarily selected value provided that, within the range $-N \leq m \leq N$, it
respects the condition
\begin{equation}
\label{eqn_13}
  0 \leq Q + \frac{2\pi m}{L} \leq \Re { \left[ \frac{\omega_0}{\cove} \right] }
  = \frac{\omega_0}{c_R \left( 1+ \frac{c_I^2}{c_R^2} \right) }
\end{equation}

wherein the Real and Imaginary parts of $\cove$ appear.

Replacing relation (\ref{eqn_12}) into Eq.(\ref{eqn_11}), one gets
\begin{equation}
\label{eqn_14}
  \boldsymbol{\Psi}(\rho,z,t) = e^{-i\omega_0t} e^{iQz} \sum\limits_{m=-N}^{N} A_m \;
  J_0((k_{\rho R_m} + ik_{\rho I_m})\rho) \; e^{-\beta_{I_m} z} \, e^{i\frac{2\pi m}{L}z}
\end{equation}

\noindent  To find out the limiting values of the
attenuation term $e^{-\beta_{I_m} z}=e^{-\alpha \cos(\theta_m) z}$, let us replace expression (\ref{eqn_12}) of $\beta_{R_m}$ into
Eqs.(\ref{eqn_9}) and get
\begin{equation}
\label{eqn_15}
  \beta_{I_m} = \left( Q + \frac{2\pi m}{L} \right) \frac{\alpha}{k}
\end{equation}

\noindent The minimum, maximum and central values of $\beta_{I}$ are then given by
\begin{align}
\label{eqn_16}
 (\beta_I)_{\text{min}} &= \left( Q - \frac{2\pi N}{L} \right) \frac{\alpha}{k} \nonumber \\
 \nonumber \\
 (\beta_I)_{\text{max}} &= \left( Q + \frac{2\pi N}{L} \right) \frac{\alpha}{k} \\
 \ \nonumber\\
 (\beta_I)_{m=0} &= Q \frac{\alpha}{k} \equiv \tilde{\beta}_I \; , \nonumber
\end{align}

\noindent respectively.

\noindent The spread of the limiting values of $\beta_I$ can be expressed by
\begin{equation}
\label{eqn_17}
  \Delta = \frac{(\beta_I)_{\text{max}} - (\beta_I)_{\text{min}}{\tilde{\beta}_I}}
               = 4\pi \frac{N}{LQ}
\end{equation}

\noindent
Then, if $\Delta \ll 1$ the values for the imaginary component of the longitudinal wavenumber will be approximately similar, so that \ $\exp{-\beta_{Im} z} \approx \exp{-\tilde{\beta}_{I} z}$, \ and superposition (\ref{eqn_14}) can be regarded when $\rho = 0$ as a truncated Fourier series, which can
reproduce the desired longitudinal intensity pattern $|F(z)|^2$ when choosing the coefficients
\begin{equation}
\label{eqn_18}
  A_m = \frac{1}{L} \int_0^L F(z)\; e^{\tilde{\beta}_I z} e^{-i\frac{2\pi m}{L}z} \textit{d} z
\end{equation}

As we can see in the coefficients (\ref{eqn_18}), the envelope function $F(z)$ of the FW is \textit{compensated}, since the beginning, for the effects of attenuation by the presence of the term $e^{\tilde{\beta}_I z}$, which does indeed counteract
in (\ref{eqn_14}) the effects of $e^{-\beta_{I_m} z}$.

\

The number $2N +1$ of terms in Eq.(\ref{eqn_14}) is limited by the condition (following from Eq.(\ref{eqn_13})):

$$N \leq {{L} \over {2\pi}} \; \left[ \Re {({{\omega_0} \over {\cove}})} - Q \right] $$

depending in particular on $Q$. The maximum possible value corresponds to $Q = \omega_0 / {2 c} \equiv Q_0$ and becomes

\begin{equation} \label{eqn_19}
  N \leq N_{\rm max} = \frac{L\omega_0}{4\pi c}
\end{equation}

where for simplicity we supposed (as it often happens in the cases considered in this work) that $c_R >> c_I$, so that
$\Re {(\omega_0 / \cove)} \approx \omega_0 / c_R$ and $|\cove| \approx c_R \equiv c$. Quantity $c$, the sound speed in a water-like medium, can be assumed to be approximately 1540 m/s.
[Care should be taken, however, when really adopting the limiting value $Q = Q_0$  since it commonly leads to unpractical transducer sizes ($R>50$\,mm), with too many rings ($N_r>100$)].

\noindent The last relations can be easily seen at work in Figure \ref{fig_2}.

Another issue we have discussed elsewhere\cite{pregoI,zamboni2011}, is that lower values of $Q$ imply narrower spots for the resulting FWs. Their size can be estimated by finding out the radius of the transverse spot, as:
\begin{equation} \label{eqn_20}
  \Delta\rho = \frac{2.4}{\sqrt{\omega_0^2/c^2 - Q^2}}
\end{equation}

Let us conclude this Section by stressing that such a method allows constructing, in absorbing media, non-diffracting beams
whose longitudinal intensity pattern can be arbitrarily prefixed: In particular, one can therefore obtain beams resisting diffraction and attenuation, as we are going to see in Sec.6.

\section{Acoustical Restrictions for Ultrasonic FWs}
So far, we discussed the attenuation of ultrasound in water and the way of modeling acoustic FWs in a lossy medium. In this section our aim is discussing the restrictions imposed by Acoustics when selecting the appropriate values of the parameters for the
ultrasonic FWs. These limitations are mainly due to two causes: \ i) the low values of the ratio
${w_0} / c$, and \ ii) the value of $L$, which normally is $L<1 \;$m.

From Eq.(\ref{eqn_13}) we can notice that, whatever the values selected for the real part ($\beta_{R_m}$) of the longitudinal wavenumbers, they must remain below the limit $\Re {[{\omega_0}/{\cove}]}$. \ Then, if one tries to create a FW with a carrier frequency for example of $f_0=2\;$MHz, the above mentioned limit\footnote{Here and in the forthcoming we will approximate this limit by ${\omega_0}/{c}$, disregarding the use of  $\Re {[{\cove}]}$, which implies a negligible difference when $c_I$ (or $\alpha$) are small, like in this work.} will approximately be ${\omega_0}/{c}\simeq 8160$.

\begin{figure}[t]  
\centering
\includegraphics[width=3.5in]{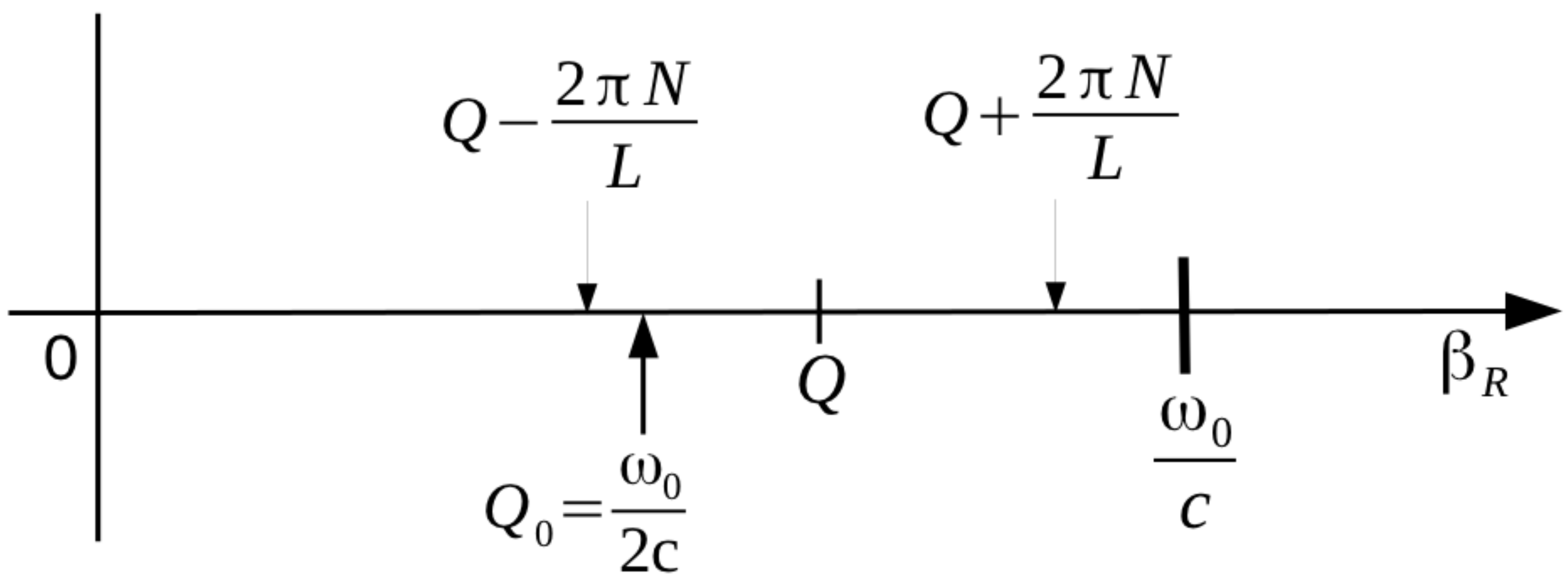}
\caption{Range and limits for the longitudinal wave numbers $\beta_{R_m}$ in the case of acoustic Frozen Waves (FW).}
\label{fig_2}
\end{figure}

\noindent Compare this value with the one corresponding to optical FWs: For example, imagine the case of an optical FW constructed via a red laser with $\lambda=632$\;nm \ and \ $c=3\,10^8$\;m/s. \ The limiting value obtained in this case is $\frac{\omega_0}{c}\approx  9.94 \times 10^6$; which is more than three orders of magnitude bigger. This is a known advantage of optical FWs with respect to their acoustical counterparts, because it allows a very large increase of the parameter $N^{\uparrow}$, for instance $N\to 100$, which in turn does enormously enhance the details of the resulting FWs. \  By contrast, in the case of ultrasonic FWs the lower value of the ratio ${w_0}/{c}$ imposes a reduction of the maximum value for the longitudinal wavenumber, \ $(\beta_{R})_{\text{max}}= Q+{2\pi N}/{L}$ \ (see Fig.\ref{fig_2}). As already noticed, this maximum value depends also on $L$, and on the used value of $N$, which controls the number of the Bessel beams entering Eq.(\ref{eqn_14}).

A possible solution to this problem would be increasing the operating frequency $f_0^{\uparrow}$, which linearly raises the mentioned limit. However, this has to be done with care, because it may require a re-dimensioning of the
transducer rings (width and kerf), which should correspondingly be decreased [This may result in technological problems if  the dimensions are too small, and even in a distortion of the FW pattern when such dimensions do not meet the required  constraints (cf. Ref.\cite{pregoI})].\\

As we mentioned earlier, the range of the longitudinal wave numbers $(Q-\frac{2\pi N}{L} ; \ Q+\frac{2\pi N}{L})$ can be located anywhere inside the interval $(0;{\omega_0}/{c})$; however the maximum excursion is obtained when $Q=Q_0 \equiv {\omega_0}/{2c}$ (see again Fig.\ref{fig_2}).

\begin{figure}[t]  
\centering
\includegraphics[width=3.5in]{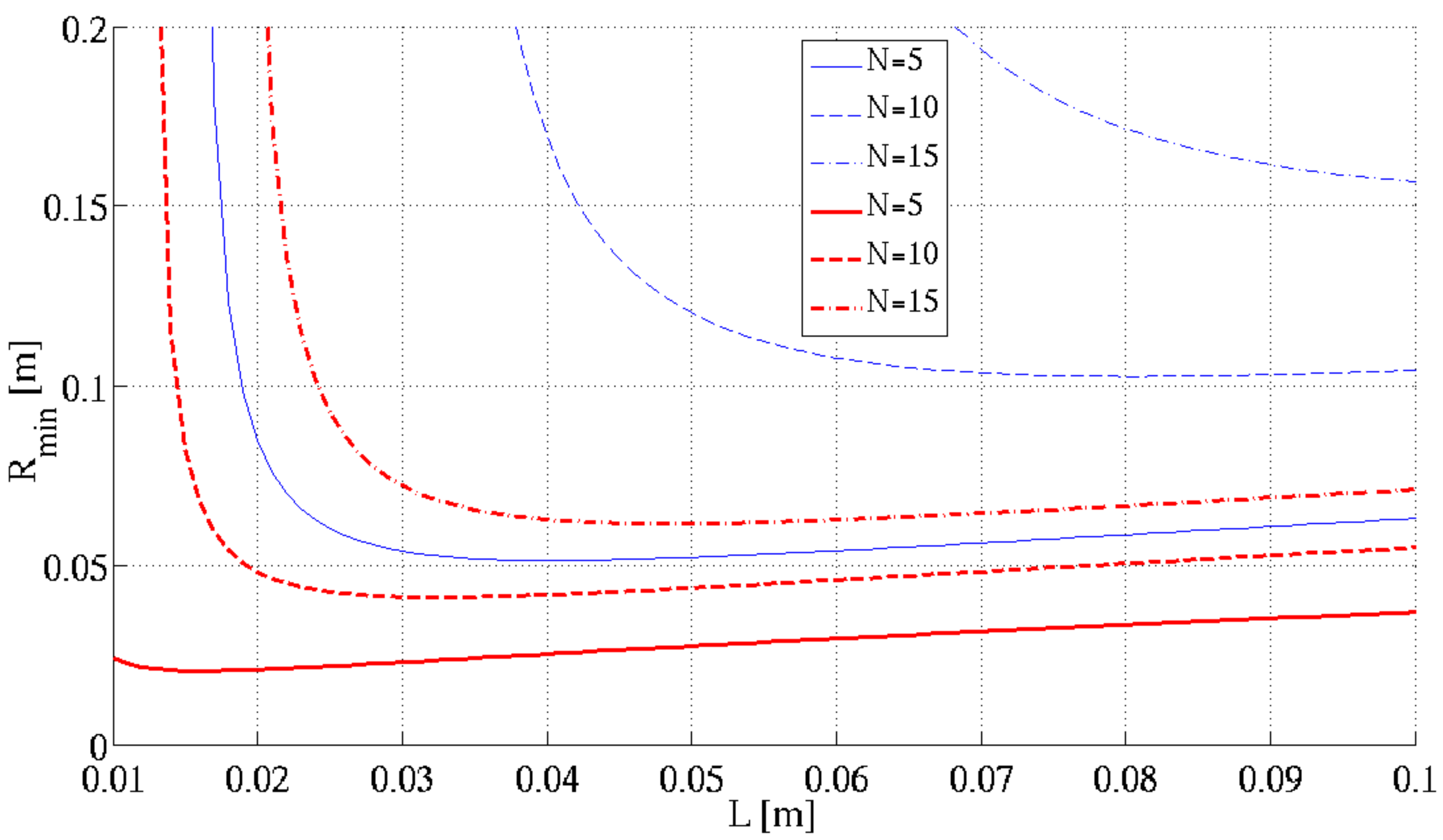}
\caption{Relationship between the aperture radius $R_{\text{min}}$ and the chosen interval $L$, for different values of the parameter
$N$. \ The thinner lines correspond to $f_0=1$\;MHz, while the thicker ones to $f_0=2.5$\;MHz.}
\label{fig_3}
\end{figure}

The terms ${2\pi}/{L}$ and $\Delta_0$ depend of course on the value of $L$. One should also notice that, whenever the frequency is increased, the ratio ${\omega_0}/{c}$ gets linearly augmented: This may allow increasing the detail of the FWs by adding more terms in superposition (\ref{eqn_14}). This can result to be useful for example when constructing energy spots highly localized along the $z$ axis. \ In the same way, when the frequency is lower, the values for $\Delta_0$ get higher, approaching unity [this might even violate, however, our assumption about the similarity of the imaginary parts of the longitudinal wavenumbers, i.e, $\beta_{I_m} \approx \tilde{\beta}_I$].

Up to now, we have discussed only the role of the ratio ${\omega_0}/{c}$ as a limiting factor for the FW creation.
A second restriction for acoustic FWs comes from the fact that $L$, in most cases, is in the range of the centimeters, i.e., $L<1\;$m: This makes the quotient ${2\pi N}/{L}$  increase.
One possible solution to this problem is augmenting $L^\uparrow$ as much as possible, so that the mentioned quotient decays sufficiently; and afterward trying to add more Bessel beams into Eq.(\ref{eqn_14}). The negative side of this option is that it might require a bigger transducer for the generation of the same FW.

We can summarize our discussion about the FW parameter restrictions, by collecting them into one expression for the determination of the minimum required radius\footnote{An alternative expression can be found also in Eq.(19) of Ref.\cite{zamboni2005}.}
$R_{\rm min}$, a sufficient condition for it being


\begin{equation} \label{eqn_21}
  R_{\text{min}} = L \sqrt{\frac{w_0^2}{c^2 \beta^2_{R \ m=-N}} -1}
\end{equation}

\noindent The \textit{sufficient} (but not necessary) condition (\ref{eqn_21}) can be derived by setting $(\beta_{R})_{\text{max}}={\omega_0}/{c}$ (see Ref.\cite{pregoI}). \ To see the dependence of the emitter radius $R_{\rm min}$ on $L$ and on parameter $N$, in Fig.\ref{fig_3} we show relation (\ref{eqn_21}) at work for two frequencies: \
$f_0=1\;$MHz (dashed lines), \ and \ $f_0=2.5\;$MHz (thicker lines). \ Notice
how the size required for the radiators are in general bigger than those normally used in laboratory applications ($R_{\text{min}}>1\,\text{cm}$). One can also notice how for certain cases of $N$ and $L$, the size of the transducer becomes completely unpractical. \ The (not mutually exclusive) alternatives for solving this problem are two: \ i) rising the operating frequency $f_0^{\uparrow}$ and lowering $N^{\downarrow}$ for the FWs, or \ ii) disregarding the rather conservative approach represented by Eq.(\ref{eqn_21}), and playing with different interval widths $L$ during the construction of the FWs:  We shall deal with this question in the Appendix.

\section{Method for the Calculation of Ultrasonic Fields}

\noindent In this section we shall summarize the technique employed for the calculation of the ultrasonic FWs fields in water, including the attenuation effect. \ To this aim, we use the well known impulse response (IR) method\cite{stepanishen1971,jensen1992}, in conjunction with a linear model of the medium attenuation\cite{jensen1993}.

The basis of the IR method is the linear system theory, which allows the separation of the spatial and temporal features of the acoustic field. Then, the IR function can be derived directly from the Rayleigh integral\cite{harris1981,goodman2005} by the expression:
\begin{equation}
\label{eqn_22}
  h(\rbf_1,t) = \int_{S} \
  \frac{\delta \left( t-\frac{|\rbf_1-\rbf_0|}{c} \right) } {2\pi |\rbf_1-\rbf_0|} dS \ .
\end{equation}

\noindent Equation (\ref{eqn_22}) assumes the emission aperture $S$ to be mounted on an infinitely rigid baffle, with $\rbf_0$ denoting the location of the radiator, and $\rbf_1$ designating the position of the considerer field point. The speed of sound in the medium is symbolized by $c$, while $t$ denotes the time variable. \ This expression is nothing but the statement of Huyghens' principle, and it allows computing the acoustic field\footnote{This is the differential field pressure, relative to the static atmospheric pressure $P_0$.} by adding up all the spherical wave contributions from the small elements that constitute the radiating aperture.

Interestingly, it is also possible to reach the same results by using the acoustic reciprocity principle. This procedure builds the $h$ function, for a particular field point $P$, by finding the angular widths (in radians) of the curve-arcs which result from the intersection of a spherical wave emanating from $P$ with the surface $S$ of the acoustic radiator. \ In both cases, the result is that the $h$ function depends on both the form of the emitting element, and its relative position with respect to the point where the acoustic field is being calculated.

\hyphenation{atte-nua-tion}
Although the original IR formulation assumes a flat, or gently curved, aperture\footnote{The aperture dimensions must be large compared to the field wavelength.}, radiating into an homogeneous medium with no attenuation, nevertheless the effects of the medium absorption can be included by properly modifying the IR function $h$. This can be performed by applying to the Fourier transform response $H(f)$ a ``filtering" function $A(f)$ which accounts for the medium attenuation,
and can be chosen as a linear model\cite{jensen1993} of the attenuation phenomenon. The attenuated IR function will be therefore:
\begin{equation}
\label{eqn_23}
h_a(x,y,z,t) = \mathcal{F}^{-1}\lbrace A(f)H(f)\rbrace .
\end{equation}

\noindent On using the modified IR function, the acoustic pressure can be obtained as
\begin{equation}
\label{eqn_24}
p(x,y,z,t) = \rho_{m} \frac{\partial v(t)}{\partial t}*h_a(x,y,z,t) \; ,
\end{equation}

\noindent where the symbol $*$ denotes the time convolution of the modified radiator impulse response function $h_a$ with the time derivative of its surface signal velocity $v$, while $\rho_{m}$ accounts for the density of the medium.

As we discussed in Section 2, the looses in the medium can be described by assuming a linear lossy equation; which introduces the attenuation by means of the parameter $\alpha$ expressed in Np/m.\\

\begin{figure}[t]  
\centering
\includegraphics[width=1.5in]{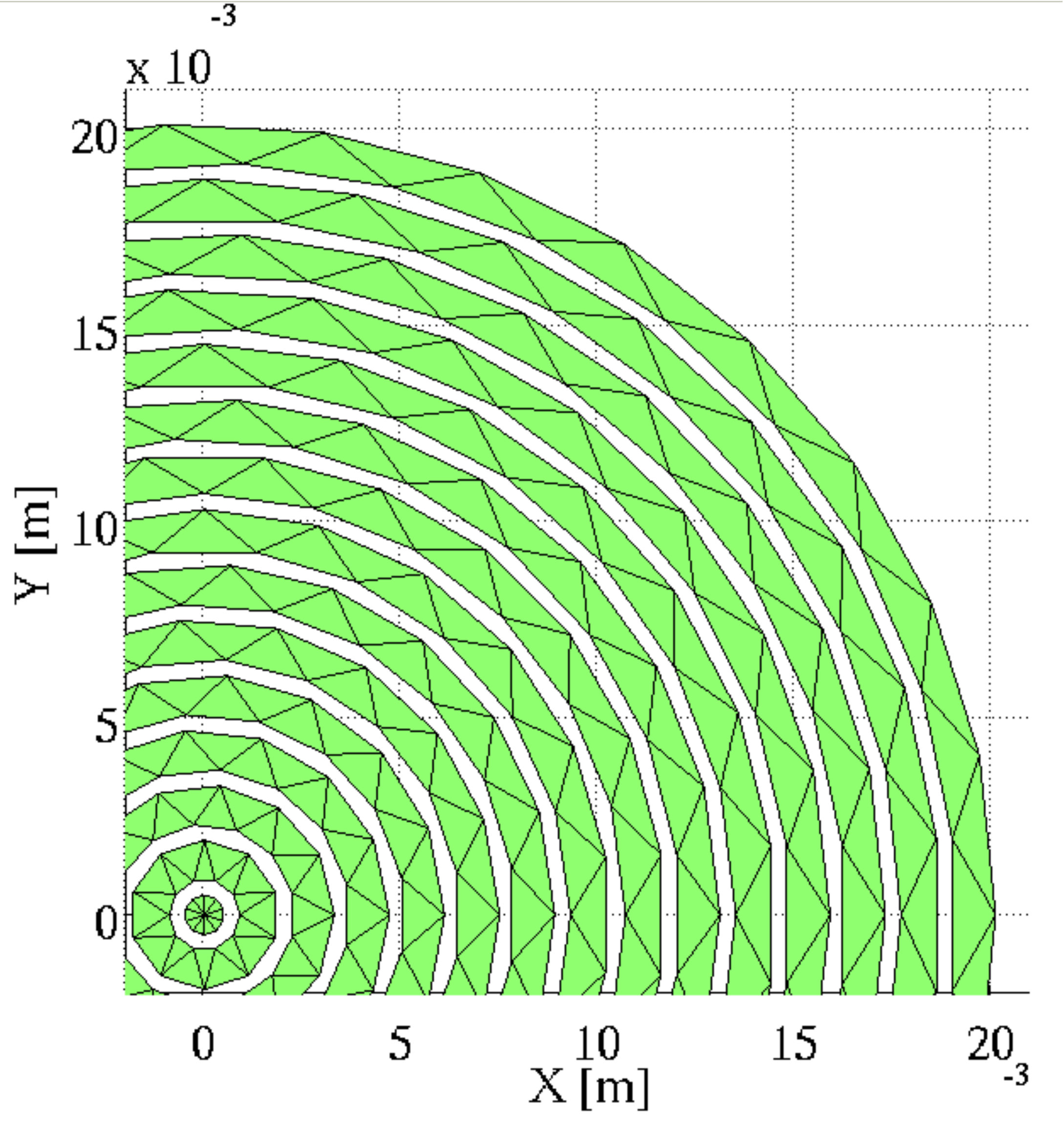}
\caption{Example of annular radiator, created by the ``Field II toolbox" package, with radius $R=20\;$mm, and with ``$N_r=15$ rings composed by $628$ triangular elements (see the text).}
\label{fig_4}
\end{figure}

In order to simulate our acoustic FWs in an environment more closely resembling the human body, we set the $\alpha$ parameter in our model to values typically found for human tissues\cite{jensen1993,rossing2007}, i.e., $0.7\le a \le 1.7 \;$ dB/(cm \ MHz), with $a=8.686\,\alpha$. \ Although it is true that this simplification may not account for all the processes occurring in reality, we consider this approach to constitute a reasonable approximation.

It is also important to stress that during the simulations, although the absorption of the wave energy will continue to occur in a normal fashion, the superposed beams will be however able to reconstruct themselves, due to the nature of their transverse field distribution and to the introduction of the compensating factor $e^{\overline{\beta}_I z}$ in the coefficients of Eq.(\ref{eqn_18}).\\

We have implemented the above method into a Matlab program employing the {\em Field II toolbox}\cite{jensen1992,jensen2010}; which allows the introduction of the medium attenuation by means of linear model based on $\alpha$. \ The annular radiators were simulated by discretizing them into triangular elements (see Fig.\ref{fig_4}), which are available in the Field II package for the calculation of the $h_a(x,y,z,t)$ function.

The steps followed by the software for the computation of the acoustic FWs fields are summarized as follows:

\begin{enumerate}
\item Compute the theoretical FW pattern, $\boldsymbol{\Psi}(\rho,z)$, using the chosen parameters.
\item Determine the radiator dimensions: radius ($R$), width ($d$) and kerf ($\Delta_d$).
\item ``Sample" amplitude and phase of the FW function at the aperture location, i.e., of the complex function
$\boldsymbol{\Psi}(\rho,z=0)$.
\item Assign a sinusoidal signal $v_k$ for the ring $n_k$, using sampled values.
\item Sweep the field points ($x_{i,j};y_{i,j};z_{i,j}$) for the transducer ring $n_k$.
\item At point $P_{(i,j)}$, calculate $h_a$  and $p_{(i,j)} = \rho_{m} \dot{v}_k*h_a$.
\item Accumulate pressure for ring $n_k$ ($p=p+p_k$), and go back to the fourth step.
\item Store and display the $p$ results.
\end{enumerate}

\section{FWs in Absorbing Media: Results of Simulated Experiments}
\hyphenation{ope-ra-ting}
In this section we like to demonstrate the possibilities offered by the acoustic frozen waves, by presenting three different examples of FWs operating in a water-like medium, including the effect of the absorption. Attenuation factors between $70\le \alpha \le 170\,\text{dB/m}$ have been tested at $f_0=1\,\text{MHz}$, with interval widths $L$ in the range $120\le L\le 240\,\text{mm}$.

Before going on, let us point out that the cases examined here were selected as rather general, in order to emphasize the capabilities offer by the method. \ We also chose a wide range of shapes for the envelope function $F(z)$, possibly usable in practical applications. In other words, our intention has been that of testing general enough patterns, rather than some particular applications, also in benefit, a priori, of both usefulness and simplicity.

\hyphenation{pro-pa-ga-tion}
As we mentioned in the Introduction, because of the use of zero order Bessel beams we expect a high degree of transverse localization too during propagation, besides the beam resistance to diffraction and absorption. As we know, such good properties are due to the self-reconstruction capability of the Bessel beams, as well as to the addition of our compensating term \
$e^{\tilde{\beta}_I z} \approx e^{i\alpha\cos(\theta)z}$ in Eq.\ref{eqn_18} \ (which counteracts the attenuation effects\cite{zamboni2006,zamboni2011}).

All of the \textit{simulated} FW patterns have been obtained by using \textit{the same} ultrasonic radiator, with radius $R\simeq 31\,$mm and $N_r=35$ rings, each one endowed with a width $d=0.6\;$mm and a kerf of $\Delta_d=0.3\,$mm. Al already mentioned,
the simulated FWs have been moreover produced by exploring the effects of different values of $L$, as so as to improve our results for the fields in the presence of absorption.

The speed of sound inserted in the program is $c \simeq 1540\,$m/s, and the operating frequency has been fixed in all cases at $f_0=1\,$MHz, which corresponds to a wavelength of $\lambda\simeq 1.54\;$mm.
The only parameters we varied during the simulations (apart from the chosen envelope $F(z)$ itself, of course) are the medium attenuation coefficient $\alpha$, expressed e.g. in Np/m [sometimes replaced by a value in dB/(cm \  MHz)], the value $L$ chosen for the FW, and the value $N$ which determines the ($2N+1$) number of Bessel beams in Eq.(\ref{eqn_14}).

The corresponding couple of Figures, shown in this Section (namely, Figs.8-9, 11-12, and 14-15) depict: First, the theoretical FW pattern obtained in the ideal case of an infinite aperture [i.e., the 3D plot of the function $\boldsymbol{\Psi}(\rho,z)$]; and, Second, the result of our Impulse Response (IR) simulation of its experimental generation by a finite aperture, respectively. The sampling frequency used in the IR method has been set to $f_s=100\;$MHz.

The details in the corresponding spatial grids ($\rho,z$) had in practice to be reduced, with respect to (w.r.t.) the ideal patterns. Then, suitable intervals of $\Delta z=0.4\,$mm and $\Delta \rho=0.1\,$mm have been selected for our {\em Field II} simulated plots.
Also, due to the adoption of colors (present online only) for the graphs, light-effects were added in order to enhance the detail visibility also on paper.

\subsection{Case 1}

\noindent Our first choice of acoustic FW to be generated consists in a uniform envelope $F(z)$, with $L=60\;$mm. This FW is defined in Eq.(\ref{eqn_25}) as existing  between $l_1={L}/{10}=6\,\text{mm}$,\ and \ $l_2={9L}/{10}=54\,\text{mm}$, with $N=6$: That is, $2N+1=13$ Bessel beams are superposed. The interval width is chosen to be $L=120\,\text{mm}$, and the attenuation of the medium has been set either to $0.7\;$dB/(cm \ MHz) or to $70\;$dB/(cm  \ MHz), at $f_0=1\,\text{MHz}$:

\begin{equation}\label{eqn_25}
F_1(z)= \begin{cases}
      1 \ &\text{for $l_1 \leq z \leq l_2$}\\
      0 \ &\text{elsewhere}\\
    \end{cases}
\end{equation}

\begin{figure}[h]  
\vspace{-3mm}\centering
\centering
\includegraphics[width=3in]{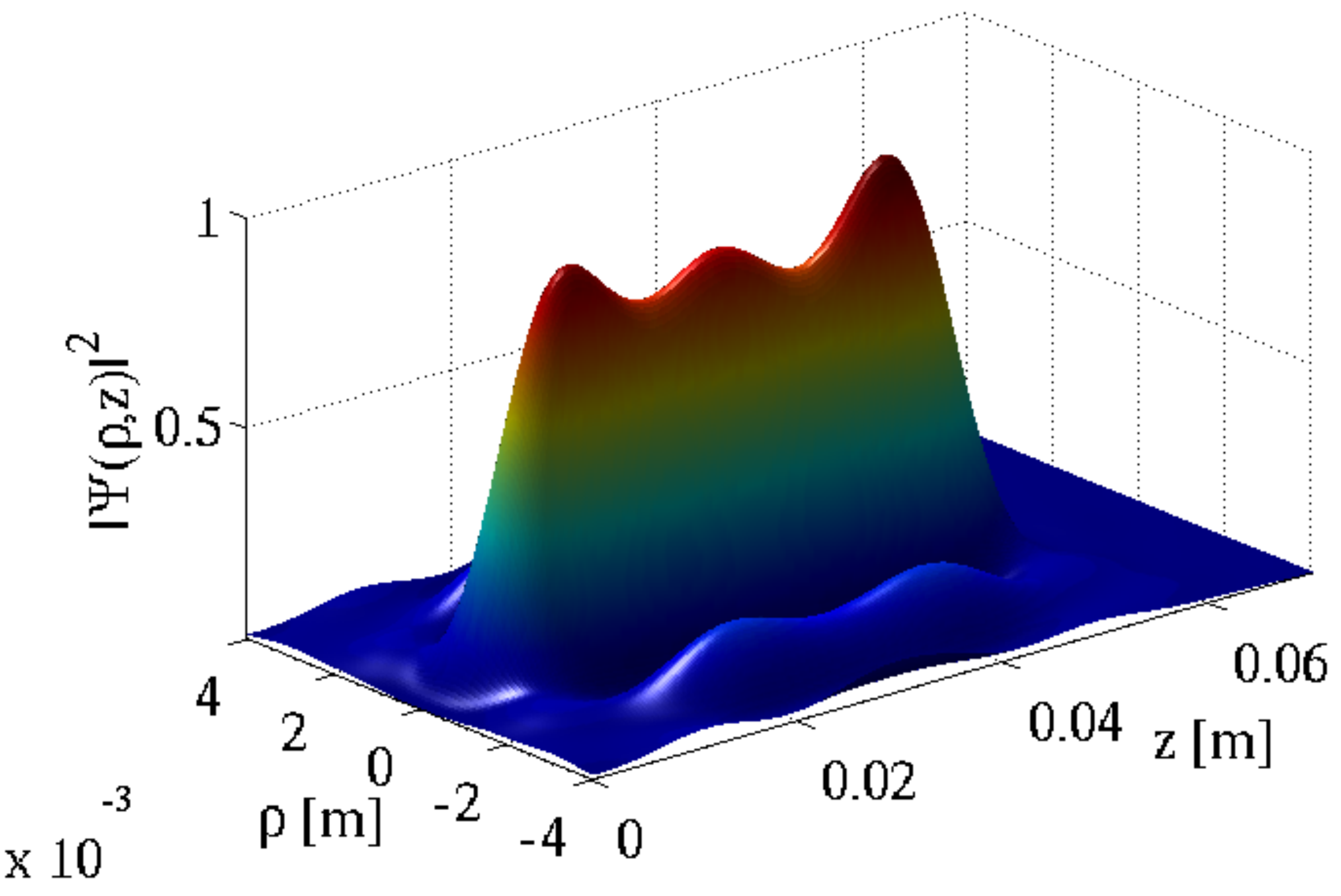}
\caption{Theoretical FW chosen in the Case 1 of Sec.6. Settings:
$L=120\,\text{mm}$; $N=6$; attenuation of the medium $0.7\;$dB/(cm
\ MHz), and $f_0=1\,$MHz.} \label{fig_8}
\end{figure}

\begin{figure}[h!]  
\centering
\includegraphics[width=3in]{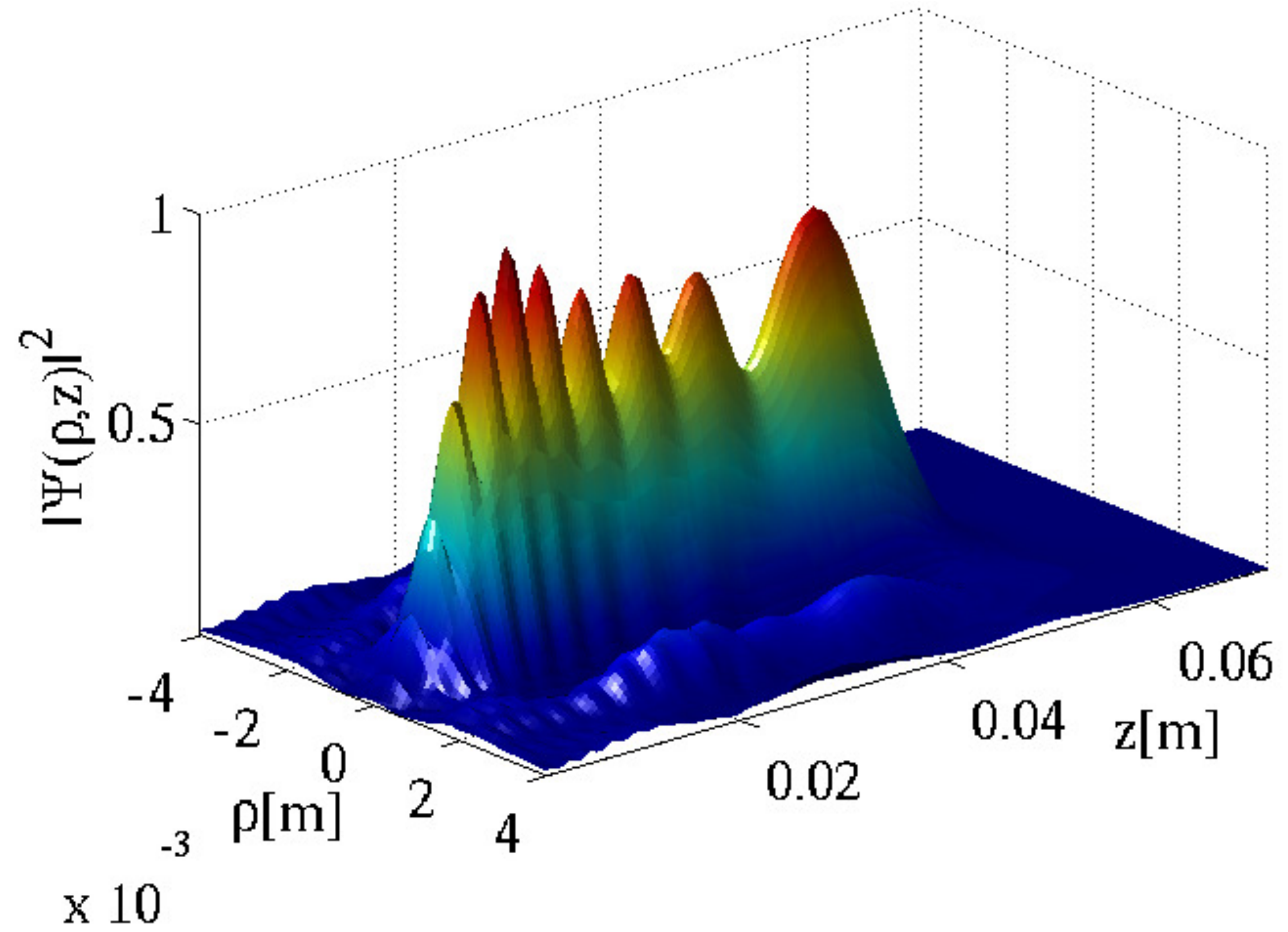}
\caption{Simulation of the experimental production, by the IR Method, of the FW in the previous figure, that is, for Case 1 of Sec.6. The settings are the same as before, except that now we are using a $N_r=35$ rings annular radiator, with radius $R \simeq 31\,$mm.}
\label{fig_9}
\end{figure}

\begin{figure}[t]  
\centering
\includegraphics[width=3.4in]{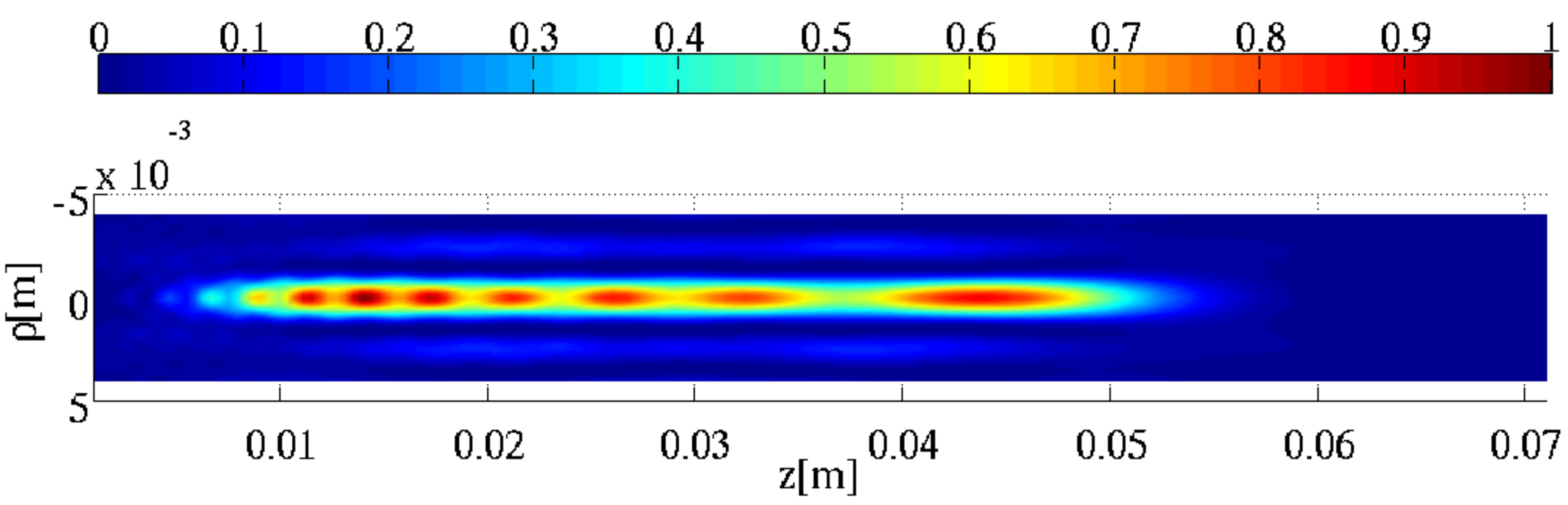}
\caption{Top view of the simulated intensity pattern in the previous Fig.\ref{fig_9}, corresponding to Case 1 of Sec.6. The size of the spot achieved is $2\Delta\rho_1 \backsimeq 3$\,mm.}
\label{fig_10}
\end{figure}

A plot of the theoretical frozen wave is shown in figure \ref{fig_8}, while the results from the impulse response method are presented in Figs.\ref{fig_9} and \ref{fig_10}. The last plot shows a (properly scaled) top view of Fig.\ref{fig_9}, showing the achieved degrees of axial and longitudinal localization. The size of the FW spot is approximately $2\Delta\rho_1 \backsimeq 3$\,mm. A rounded off value of the complex speed of sound in the medium, plus some parameters of the FW, are shown in Table \ref{tab_3}.

\begin{table}[h]
\caption{Parameters for the FW corresponding to Case 1 of Sec.6}
\hypertarget{tab1}{\label{tab_3}}
\small{\footnotesize{
\begin{tabularx}{9.0cm}{|c|c|c|c|c|c|}\hline
\begin{tabular}{c} $\cove$\\ $[\text{m/s}]$ \end{tabular} &
\begin{tabular}{c} $\theta_{\rm min}$\\ $[\text{deg}]$ \end{tabular} &
\begin{tabular}{c} $\theta_{\rm max}$\\ $[\text{deg}]$ \end{tabular} &
\begin{tabular}{c} $z_{\theta_{\rm min}}$\\ $[\text{mm}]$ \end{tabular} &
\begin{tabular}{c} $z_{\theta_{\rm max}}$\\ $[\text{mm}]$ \end{tabular} &
\begin{tabular}{c}\!\!\! ${\Delta}$\end{tabular}\\\hline
 $1539.994-i\,3.042$    & $2.6$ & $32.3$ & $693$ & $49$ & $0.17$ \\\hline
\end{tabularx}
}}
\end{table}

\subsection{Case 2}

\noindent The next case corresponds to two peaked regions, or localized spots of energy, with medium absorption 1.0 dB/(cm  \ MHz), that is, $100\,\text{dB/m}$, at $f_0=1\,\text{MHz}$. The envelope $F(z)$ is explicitly given by:

\begin{equation}\label{eqn_26}
F_2(z)= \begin{cases}
      1 \ &\text{for $l_1 \leq z \leq l_2$}\\
      1 \ \ &\text{for $l_3 \leq z \leq l_4$}\\
      0 \ \ &\text{elsewhere}\\
        \end{cases}
\end{equation}

\noindent with the limits: $l_1={1.5L}/{10}=10.5\,\text{mm}$, \ $l_2={3L}/{10}=21\,\text{mm}$, \ $l_3={7.5L}/{10}=52.5\,\text{mm}$, and \ $l_4={9L}/{10}=63\,\text{mm}$. The interval $L$ in this case is $L=190\;$mm, with $N=7$.

\begin{figure}[t]  
\centering
\includegraphics[width=3in]{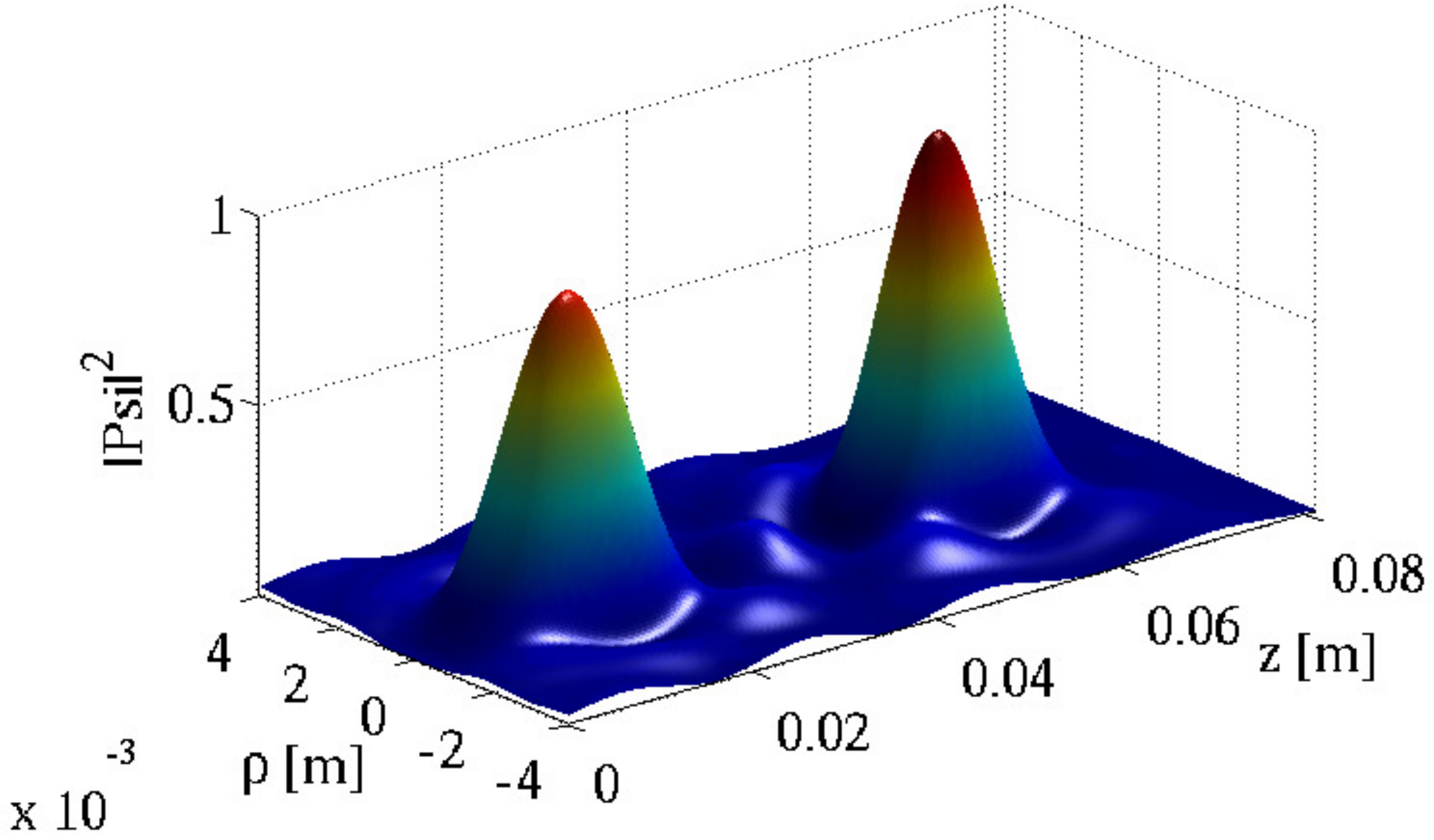}
\caption{The theoretical FW chosen in Case 2 of Sec.6. \ Settings: $L=190\,\text{mm}$; $N=7$; medium attenuation $\text{dB/(cm  \   MHz)}$, and $f_0=1\,$MHz.}
\label{fig_11}
\end{figure}

\begin{figure}[h!]  
\centering
\includegraphics[width=3in]{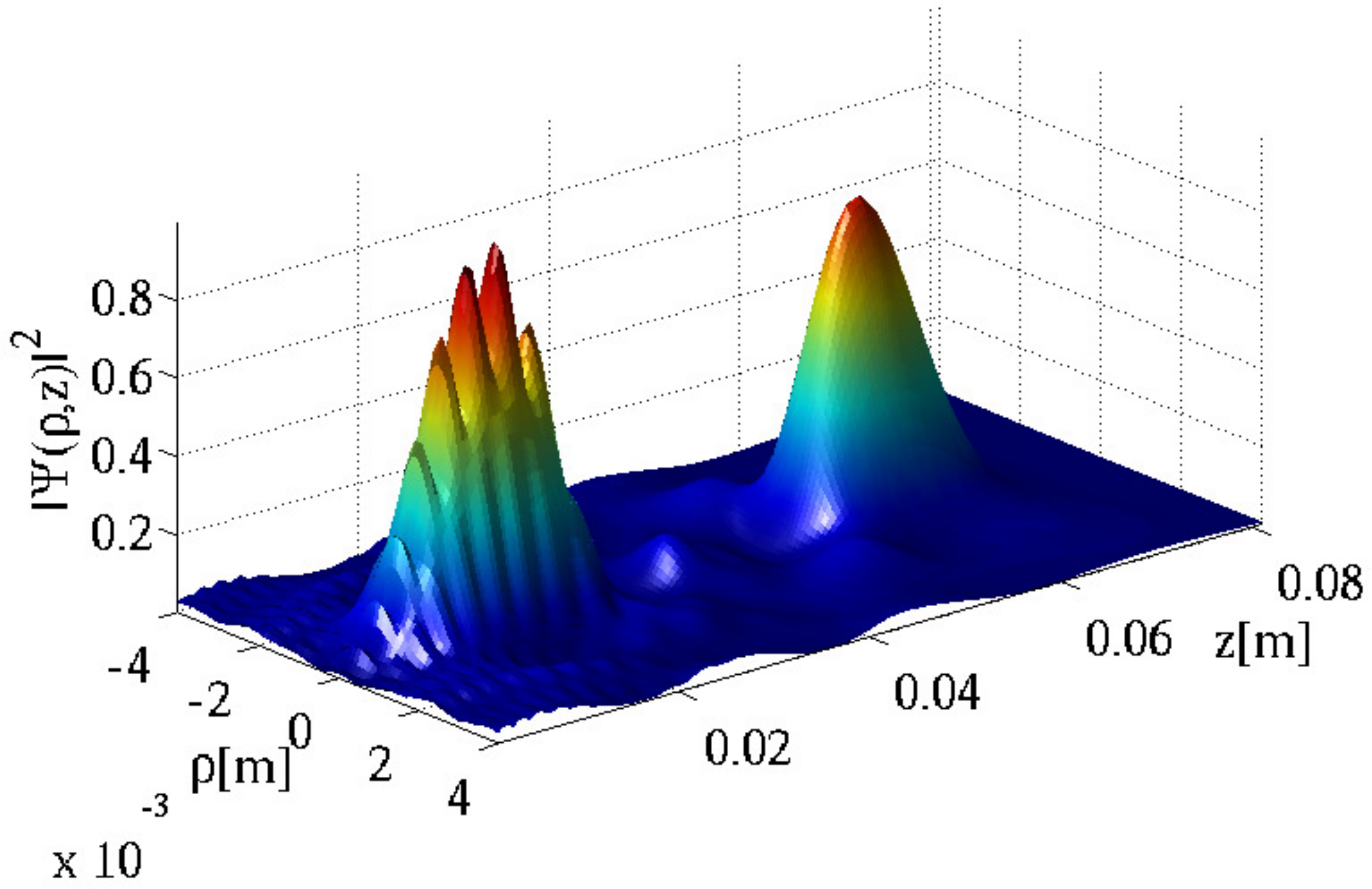}
\caption{Simulation, by the IR method, of the experimental production of the FW in the previous Figure (Case 2 of Sec.6). The settings are the same, except that we are now using an annular radiator with $N_r=35$ rings and
radius $R \simeq 31\,$mm.}
\label{fig_12}
\end{figure}

\begin{figure}[h!]  
\centering
\includegraphics[width=4in]{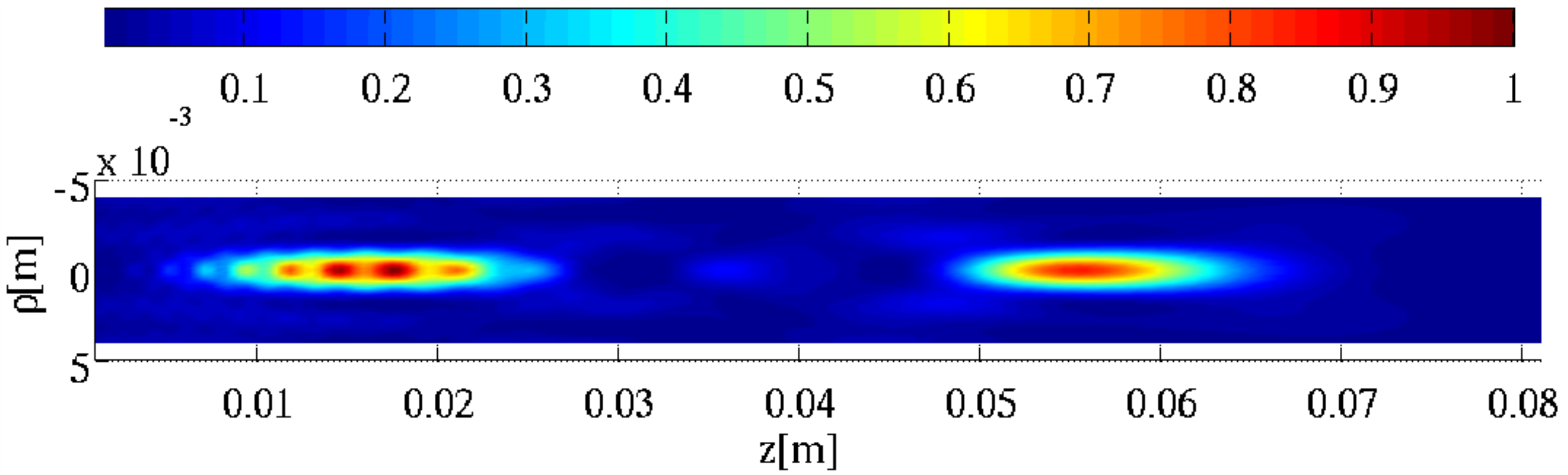}
\caption{Top (scaled) view of the FW simulated in Fig.\ref{fig_12}, corresponding to Case 2 of Sec.6. The achieved axial size of its spot is $2\Delta\rho_2\backsimeq 3.5$\,mm.}
\label{fig_13}
\end{figure}

\hyphenation{the-o-re-ti-cal}
The plots in figures \ref{fig_11} and \ref{fig_12} correspond, once again: First, to the theoretical FW, and: Second, to the {\em Field II} simulation of the generating experiment, respectively. \ Table\ref{tab_4} shows the parameters of this FW, besides the approximate value of the complex velocity of sound in the medium. \ Figure \ref{fig_13}, with regard to the simulation in
Fig.\ref{fig_12}, presents a (scaled) top view of the location of its energy peaks, which possess an axial spot size of approximately $2\Delta\rho_2 \backsimeq 3.5$\,mm.

\begin{table}[h!]
\caption{Parameters for the FW corresponding to Case 2 of Sec.6}
\hypertarget{tab2}{\label{tab_4}}
\footnotesize{
\begin{tabularx}{9.94cm}{|c|c|c|c|c|c|}\hline
\begin{tabular}{c} $\cove$\\ $[\text{m/s}]$  \end{tabular} &
\begin{tabular}{c} $\theta_{\rm min}$\\ $[\text{deg}]$ \end{tabular} &
\begin{tabular}{c} $\theta_{\rm max}$\\ $[\text{deg}]$ \end{tabular} &
\begin{tabular}{c} $z_{\theta_{\rm min}}$\\ $[\text{mm}]$ \end{tabular} &
\begin{tabular}{c} $z_{\theta_{\rm max}}$\\ $[\text{mm}]$ \end{tabular} &
\begin{tabular}{c}\!\!\! ${\Delta}$\end{tabular}\\\hline
 $1539.988-i\,4.345$    & $2.6$ & $27.7$ & $693$ & $59$ & $0.12$ \\\hline
\end{tabularx}
}
\end{table}

\subsection{Case 3}
\hyphenation{ope-ra-ting}
In this last example, we choose a growing intensity pattern for the FW, operating once more at $f_0=1\,$MHz. The absorption of the medium has been made even stronger, with a parameter $a=1.7\;$dB/(cm \  MHz). The FW envelope is given by a polynomial function [see Eq.(\ref{eqn_27}) below] corresponding in this case to
the interval $L=240\;$mm, with $N=7$.

\begin{equation}\label{eqn_27}
F_2(z)= \begin{cases}
      \frac{z^2}{2}+3z+0.1 \ &\text{for $0 \leq z \leq L$}\\
      0 \ \ &\text{elsewhere}\\
        \end{cases}
\end{equation}

\noindent Some of the parameters of this frozen waves are given in Table \ref{tab_5} below, together with the complex speed of sound in the medium.

Notice the depth of field achieved in this case ($z_{\theta_{\rm min}}\cong1.55\,\text{m}$) for the minimum axicon-angle Bessel beam, that is, for  $\theta_{\rm min}\cong1.15\,\text{Deg}$. \ We obtained such a value by adding another decimal digit to the constant $b=0.9998$ used for calculating $Q$, defined as $Q={b\,\omega_0}/{c}-{2\pi N}/{L}$.

\begin{figure}[h!]  
\vspace{-3mm}\centering
\centering
\includegraphics[width=3in]{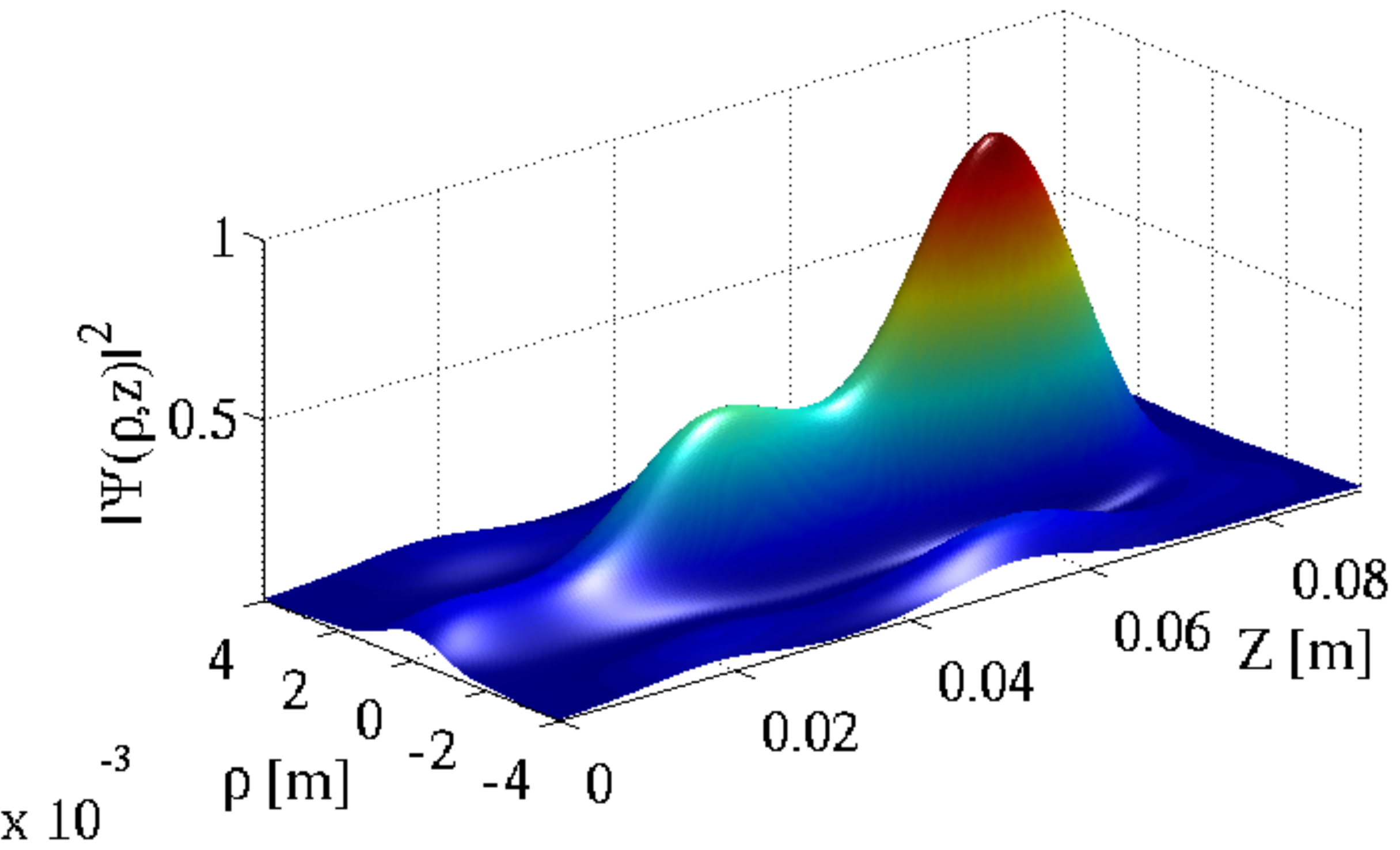}
\caption{The theoretical FW chosen in Case 3 of Sec.6. \ Settings: $L=240\,\text{mm}$; $N=7$; with medium attenuation $\text{dB/(cm  \   MHz)}$, and $f_0=1\,$MHz.}
\label{fig_14}
\end{figure}

\begin{figure}[h!]  
\vspace{-3mm}\centering
\centering
\includegraphics[width=3in]{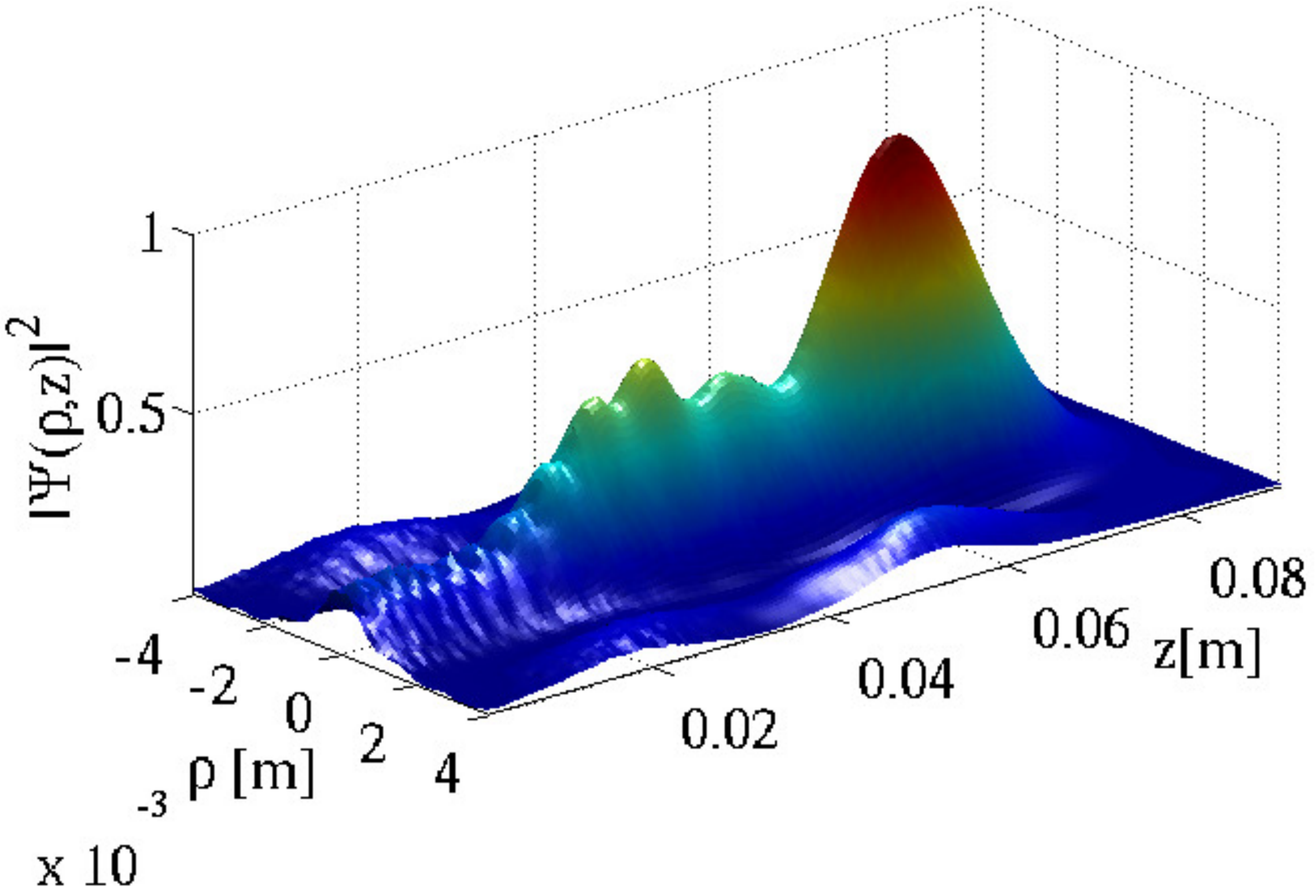}
\caption{Simulated FW corresponding to Fig.\ref{fig_14}, i.e. to Case 3 of Sec.6. \ The settings are the same, except that now we are using a $35$ rings annular radiator.}
\label{fig_15}
\end{figure}

\begin{figure}[t]  
\centering
\includegraphics[width=4in]{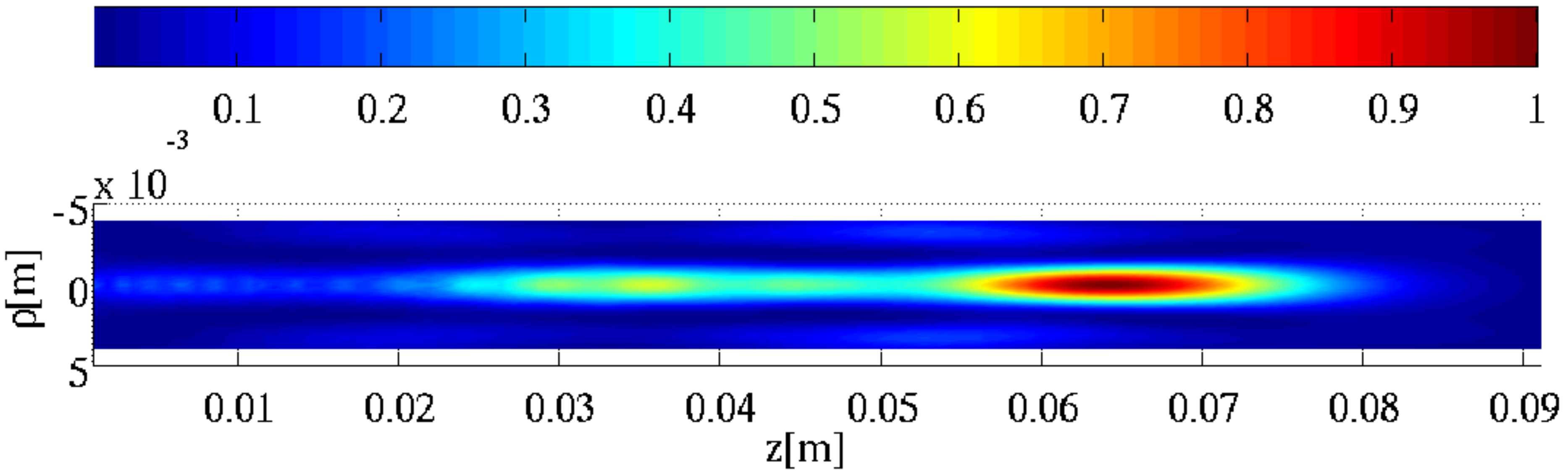}
\caption{Scaled top view of the FW simulated in Fig.\ref{fig_15}, corresponding to Case 3 of Sec.6. \ The size of the achieved axial spot is $2\Delta\rho_3 \backsimeq 4$\,mm.}
\label{fig_16}
\end{figure}

The annular aperture employed is the same as in the previous examples, that is, with $N_r=35$ rings endowed with width $d=0.6\,\text{mm}$, \ spacing $\Delta_d=0.3\,\text{mm}$, \ and radius $R=31.2\,\text{mm}$.

\begin{table}[h]
\caption{Parameters for the FW corresponding to Case 3 of Sec.6}
\hypertarget{tab3}{\label{tab_5}}
\small{\footnotesize{
\begin{tabularx}{9.0cm}{|c|c|c|c|c|c|}\hline
\begin{tabular}{c} $\cove$\\ $[\text{m/s}]$  \end{tabular} &
\begin{tabular}{c} $\theta_{\rm min}$\\ $[\text{deg}]$ \end{tabular} &
\begin{tabular}{c} $\theta_{\rm max}$\\ $[\text{deg}]$ \end{tabular} &
\begin{tabular}{c} $z_{\theta_{\rm min}}$\\ $[\text{mm}]$ \end{tabular} &
\begin{tabular}{c} $z_{\theta_{\rm max}}$\\ $[\text{mm}]$ \end{tabular} &
\begin{tabular}{c}\!\!\! ${\Delta}$\end{tabular}\\\hline
 $1539.965-i\,7.387$    & $1.15$ & $24.5$ & $1550$ & $68$ & $0.09$ \\\hline
\end{tabularx}
}}
\end{table}

\section{Conclusions}

The non-diffracting solutions to the wave equations are suitable superpositions of Bessel beams, which resist efficiently the
effects of diffraction: At least up to a certain finite distance (field-depth), when emitted by finite apertures.

\hyphenation{lo-ca-li-za-tion}
Particular superpositions of such beams allow the creation of fields $\boldsymbol{\Psi}(\rho,z)$ with a previouly chosen
intensity envelope  $|F(z)|^2$, and endowed with zero peak-velocity. We called \textit{Frozen Waves} (FW) these peculiar fields, characterized by a {\em static} envelope within which only the carrier wave propagates.

Experiments for producing acoustic FWs in ideal (non-lossy) media were already simulated in Ref.\cite{pregoI}.

On the contrary, in this paper we have focussed our attention on the creation of acoustic FWs in a (water-like) {\em absorbing}
medium.  To this purpose, we have adopted a linear model for the medium coefficient $\alpha$, which depends quadratically
on the frequency. And this model has
allowed us to express the solutions of the Helmholtz lossy equation in terms of exponentially damped plane waves.

In our approach, we compensate since the beginning such attenuation effects by the addition of the term
$e^{\tilde{\beta}_I z}$ in the Fourier coefficients' equation (\ref{eqn_18}), with the sufficient (but not necessary) condition that
$\beta_{I_m}\approx\tilde{\beta}_I$, \ or \ $\Delta \ll 1$. \
It is interesting to note that, even when the condition $\Delta \ll 1$ is not well
fulfilled, it is nevertheless possible to create FWs, successfully, in a moderate absorbing environment.

By the IR method, we simulated experiments for generating
localized fields in three different cases, all operating at
$f_0=1\,\text{MHz}$ and with same annular aperture, by adopting
increasing values of the attenuation, from 0.7 to $1.7\;$dB/(cm  \
MHz)], and various interval widths $L$.

Some problems concerning the \textit{acoustic} case of FWs have been also discussed, by pointing out, e.g., the role of the ratio ${\omega_0}/{c}$ as a factor limiting the resolution achievable for the FWs patterns. This situation
can be partially overcome by using a  higher operating frequency $f_0^{\uparrow}$. [But, since this leads of course to a reduction of the wavelength
$\lambda^{\downarrow}$, care has to be paid in order that the dimensions of the rings of the final piezoelectric
transducer remain practically realizable].

The condition that it be $L<1 \;$m was also analyzed, and different (increasing) values of $L$ considered, as so as to improve our results for the fields in the presence of absorption (cf. also the Appendix).
\ A side-effect of using large values of $L$ is, however, that it tends to increase the
amplitude of the lateral lobes: This can be observed in our Figures.\\

Among the many possible combinations of values for the FW parameters (that is, frequency $f_0$, interval $L$, number of
Bessel beams $2N+1$, and $Q$), the one that deserves more attention during the FW parameter selection
is the value of $Q$. \ This value, as mentioned before, can be anywhere inside the interval ($0,\frac{\omega_0}{c}$), as long as the corresponding  longitudinal wavenumbers $|\beta_m|$ too remain inside the same range.

An important observation is that, although the sizes of the FW axial spots decrease for lower
values of $Q^{\downarrow}$ (see Eq.\ref{eqn_20}), such values tend to produce increasing oscillations in the fields at the aperture location, i.e., for $\boldsymbol{\Psi}(\rho,z=0)$. \ And use of such resulting fields may end in a poor sampling process, which we know to have a critical importance, since it eventually determines amplitudes and phases of the sinusoidal signals that drives the rings of the radiator. \ For such a reason, we chose values as $Q={b\,\omega_0}/{c}-{2\pi N}/{L}$, with $b\to 1$, because in this manner the unwanted oscillations are reduced to a minimum, and the final frozen wave results less affected by such a phenomenon.\\

Closely related to this last point is the issue of selecting the dimensions for the annular radiator: That is, its radius
$R$, the ring widths $d$, and the spacing (or kerf, $\Delta_d$) between the rings. One does not
have a priori specific criteria for this selection, apart from the ``conservative" approach of equation (\ref{eqn_21}) which suggests the minimum possible radius for the aperture.

\hyphenation{ave-ra-ge}
What we have found experimentally in our tests is the existence of {\em a relation} between the wavelength $\lambda$, chosen for the
FW, and the average distance between peaks and valleys of the field at $z=0$ [i.e., of $\boldsymbol{\Psi}(\rho,z=0)$], suggesting as expected that lower wavelengths correlate to reduced ring widths. \ On this basis, we designed a radiator ($R\approx31\,\text{mm}$, $N_r=35$ rings) whose ring dimensions ($d=0.6\,\text{mm}$ and $\Delta_d=0.3\,\text{mm}$) are sufficiently smaller than the $\lambda$ used in the simulations. This ensures that the ring sizes do not introduce distortions in the FW fields, during the sampling process of the FW magnitude and phase. \ At the same time, they are big enough to reduce the number of rings, that is, of the electronic channels, to be used.

It is interesting to note that, even when the radius of the aperture is less than half of that suggested by
Eq.(\ref{eqn_21}) (e.g., for case 1, $R_{\rm min_1}=75.7\,\text{mm}$), we have still been able to reasonably produce FWs in the absorbing media. Indeed, that expression, as we know, was derived on the basis of a sufficient (but not necessary) condition.


Finally, as stated in the introduction, we wish to add a few words on the practical realization of acoustic FWs, especially
because the main focus of this paper has been addressed to the theoretical aspects of acoustic FWs in attenuated media, and to the restrictions imposed on them by Acoustics. \ When practically implementing FWs in ultrasound, the main problem is of course the availability of \textit{suitable} annular
piezoelectric transducers, and of the required electronic front-end to drive the rings. They do affect the
minimum number of electronic channels necessary to create the stationary wave fields, without distorting the originally chosen
FW pattern.

A second issue arises when dealing with the transducer itself. Indeed, the {\em Transfer Response} of each individual annulus of the radiator can introduce `distorting' effects in the amplitudes \& phases of the signals; and attention should be paid to the effects of signal attenuation and delay produced by the yet unknown electrical/mechanical transfer function of each ring of the (piezoelectric) transducer. To characterize each of the transducer annuli, one could
scan each piezoelectric ring with a hydrophone and measure, at the same time, the electric input signal and the acoustic
output pressure. Then, the individual ``transferences" for each of the rings can be derived, and suitable compensation factors can
be added into the amplitude \& phase ideal samplings, before the sinusoidal signals can actually drive the piezoelectric rings.

We believe that the practical realization of acoustic FWs can open the door to very interesting applications, in the field of medicine, as well as in other technological sectors.

\hyphenation{atten-tion}
\section{Acknowledgments}
\addcontentsline{toc}{section}{Acknowledgment}
\noindent  This work was supported by CNPq and FAPESP, Brazil (besides INFN, Italy). One of the authors (ER) acknowledges the research fellowship no. 2013/12025-8 by FAPESP, and another one (JLPB) the CNPq fellowship no. 500364/2013-3. Moreover, two of the authors (JLPB and ER) wish to thank Hugo E. Hern\'andez-Figueroa for continuous, helpful interest.  

\

\

\

\

\centerline{{\bf APPENDIX}}

\

\section{Effects of varying the interval widths $L$ on the produced acoustic FWs}

\

\hyphenation{se-cond} Along with our discussion about the existing
restrictions on the FW acoustic parameters, we have analyzed the
role of the ratio ${\omega_0}/{\cove}$, and of the values of $L$
commonly smaller than 1 meter. This last condition imposes that,
for a particular value of ${2\pi N}/{L}$, the quantity
$N^{\downarrow}$ has to be reduced as $L\to 0$, by lowering as a
consequence the resolution of the patterns. \ The same condition,
due to the increase of the parameter $\Delta={4\pi
N}/{(LQ)}\uparrow$, raises the spread of the $\beta_{Im}$ values,
so that superposition (\ref{eqn_14}) becomes more and more far
from a Fourier representation (when $\rho=0$).

\begin{figure}[t]  
\centering
\includegraphics[width=3.4in]{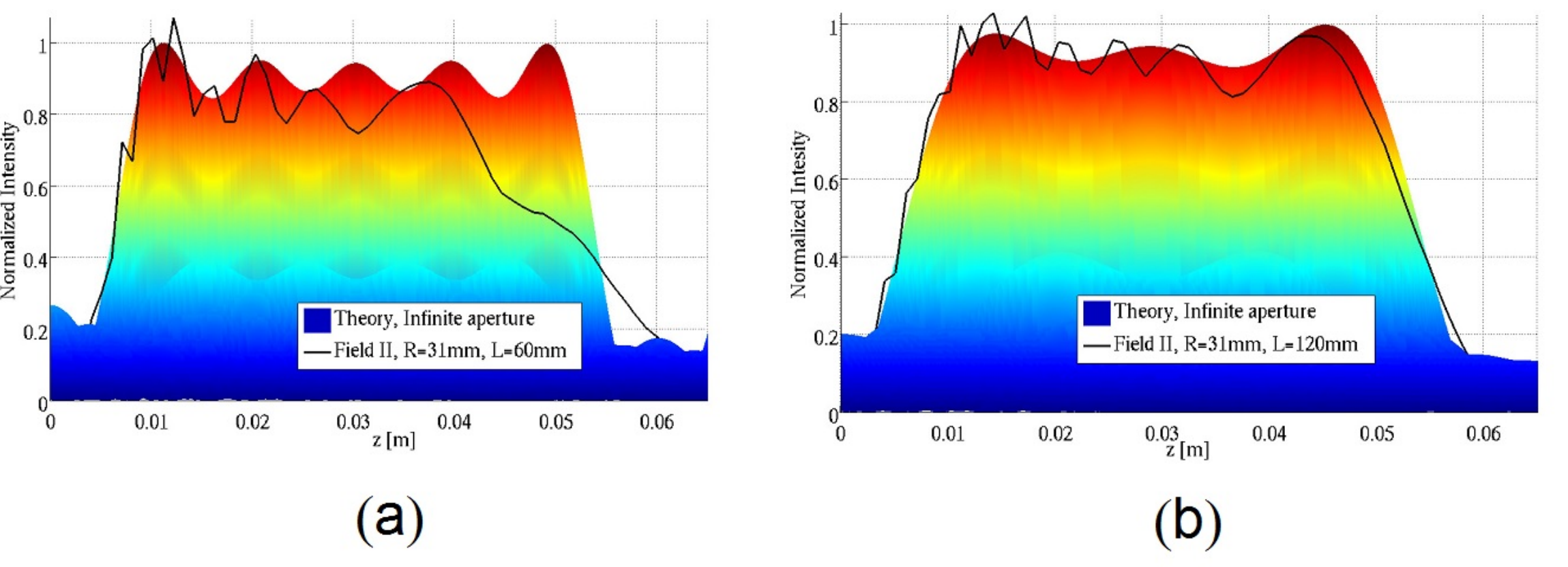}
\caption{Comparison of two {\em Field II} simulated profiles in
absorbing media, for the FW considered in Case 1 of Sec.6. \ The continuous
line corresponds to the adoption of $L=60\;\text{mm}$ in the case of Figure (a); and of a large value of $L$, namely
$L=120\;\text{mm}$, in the case of Figure (b). In both cases the emitter is the same, i.e.,
with radius $R \approx 31\;\text{mm}$ and $N_r=35$ rings, operating
at $f_0=1\;\text{MHz}$ with $N=6$, and attenuation $0.7\;$ dB /(cm
\ MHz).}
\label{fig_5}
\end{figure}

A possible way out, is having recourse to different interval-width $L$ during the computation of the acoustic FW. \
To illustrate this point, in Figs.\ref{fig_5} of this Appendix we represent the intensity profiles, along the $z$ axis, of a $f_0=1\;$MHz FW  [see Case 1 in Sec.6: Cf. Eq.(\ref{eqn_25}) therein],
simulated by us for an absorbing medium with $c \simeq 1540$, and corresponding to $L=60$ and $L=120\;$mm, respectively.

\noindent The attenuation has been set to 0.7 \; dB/(cm  \   MHz),
with $N=6$ (that is with $2N+1=13$ Bessel beams), and
$L=60\,\text{mm}$. \ Note how, when using $L=60\,\text{mm}$
(continuous line), the intensity pattern falls down before the end
of the theoretical FW (solid image); but, when increasing the
value of $L$ (dotted line), keeping the number of terms in the
expansion, the pattern gets improved.

\begin{figure}[h!]  
\centering
\includegraphics[width=2.6in]{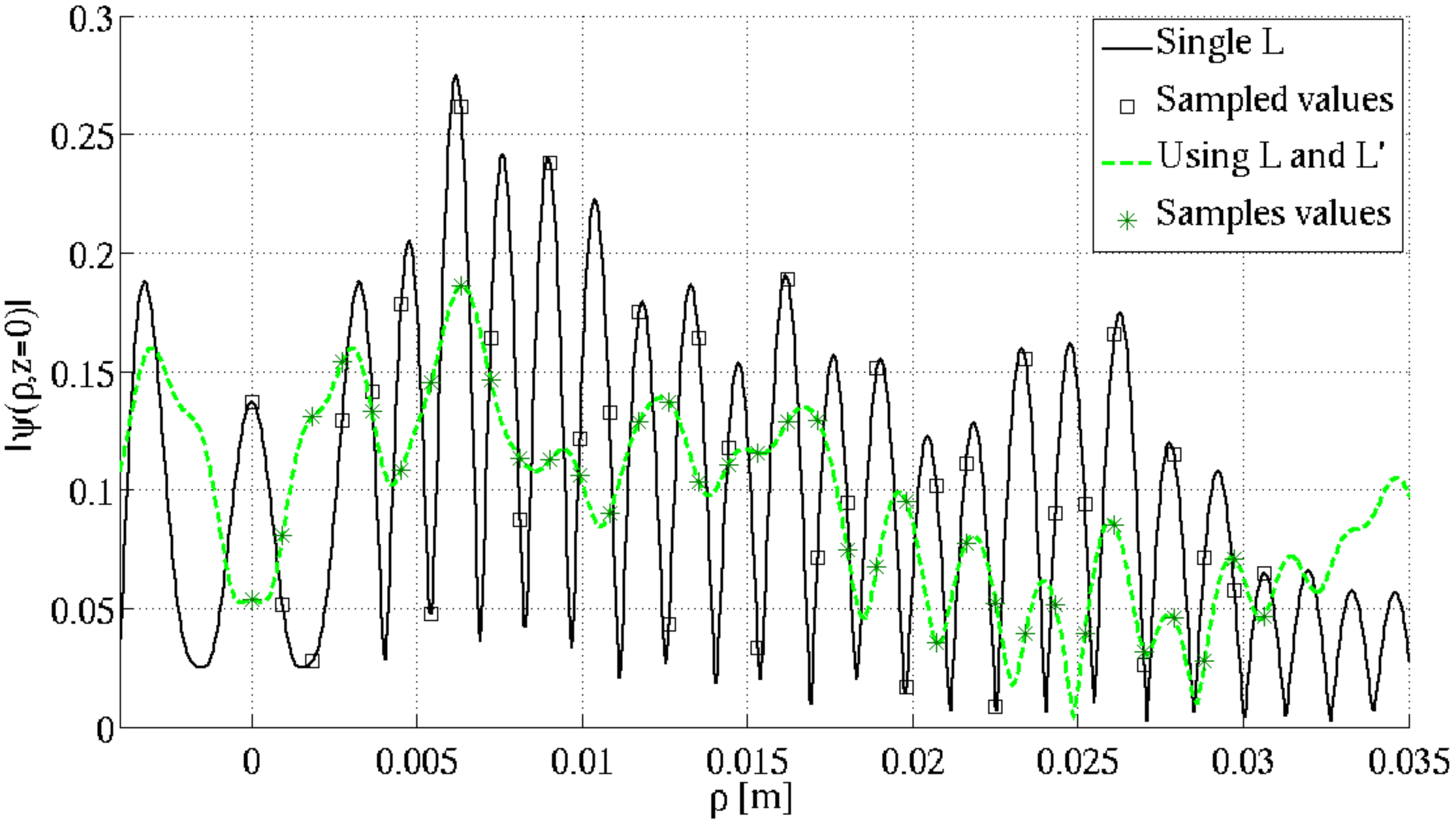}
\caption{Comparison of the FW fields at the aperture location, i.e., $|\boldsymbol{\Psi}(\rho, z=0)|$, for the profiles shown in Figs.\ref{fig_5}. The dotted line corresponds to using $L=120\,\text{mm}$; \ while the continuous line comes out when $L=60\,\text{mm}$. \ In both cases the same emitter is used, operating at $f_0=1\,\text{MHz}$ with attenuation $0.7\;$dB/(cm  \   MHz), and $N=6$.}
\label{fig_6}
\end{figure}

A second benefit of adopting larger values of $L$, is that the
fields at the aperture location, i.e, $|\boldsymbol{\Psi}(\rho,
z=0)|$, get significantly smoothed. This helps the sampling
process, that assigns the amplitudes and phases of the sinusoidal
signals, which drives the ring elements. We can observe this
phenomenon in Fig.\ref{fig_6}, which corresponds to the cases
presented in the previous figures \ref{fig_5}.

The side effect of large values of $L$ is that the original FW envelope, $F(z)$, gets distorted during the integrations in Eq.(\ref{eqn_18}). Then, a compromise has to be sought between how much we like to prevent the final
patterns from the attenuation effect, and how much we accept the original envelope $F(z)$ to be altered during the process. \ To better observe this trade-off, in Table (\ref{tab_2}) we show the values of some of the parameters
used during the construction of the above FWs profiles.

\hyphenation{ma-xi-mum}
Notice, when using for example $L=60\;$mm, how much the value of the maximum axicon angle $\theta_{\rm max}$ gets grater than when one adopts larger values for $L$. This causes the corresponding depth $z_{\theta_{\rm max}}={R}/{\tan\theta_{\rm max}}$ of the Bessel beam to be reduced, affecting the final Fourier reconstruction.

\begin{figure}[b!]  
\centering
\includegraphics[width=2.6in]{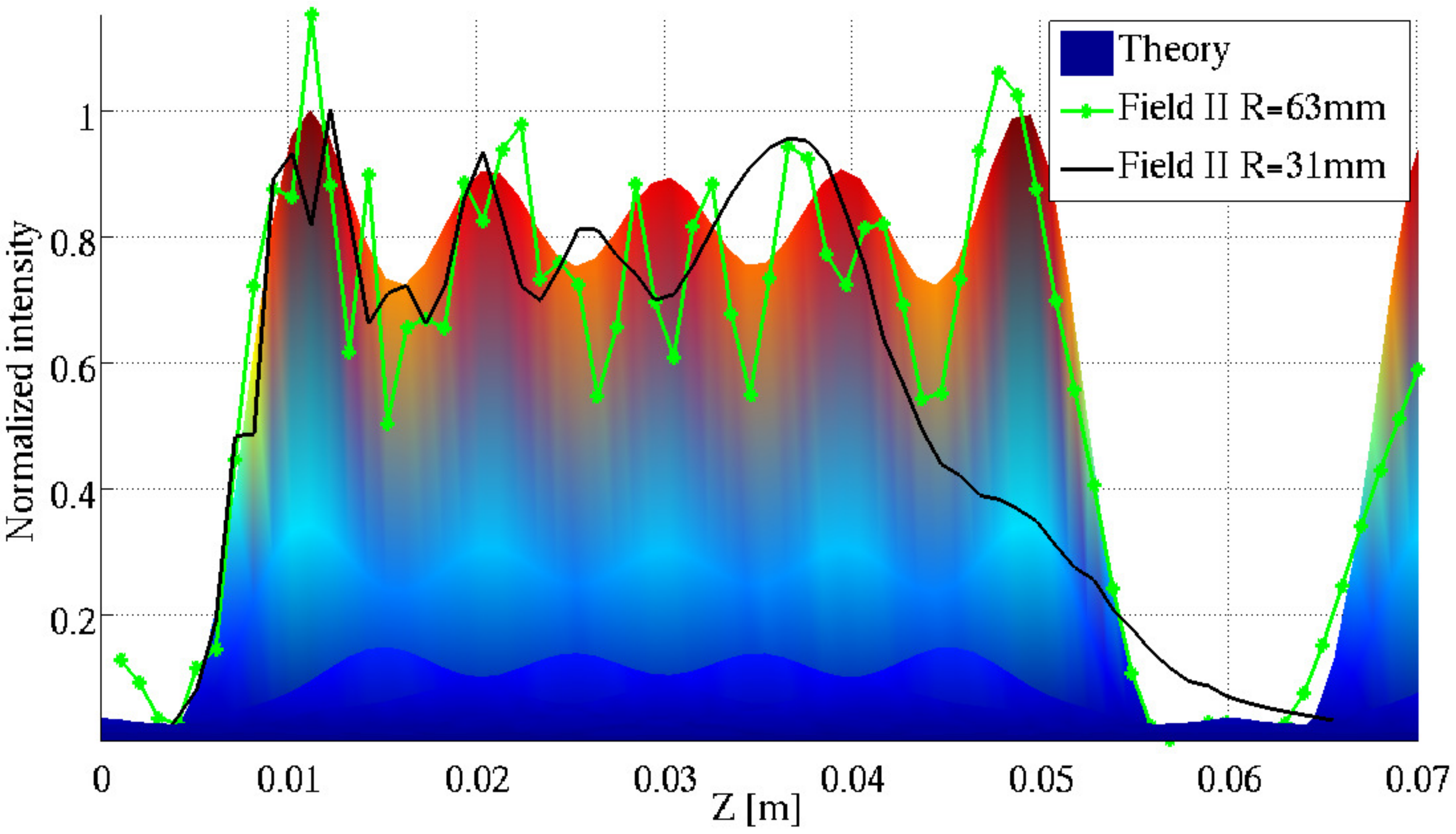}
\caption{Comparison of {\em Field II} non-attenuated (ideal media) profiles of the FW considered in Case 1, Sec.6. \ First (dotted line), using a radiator with $N_r=71$ rings and a radius of $R \simeq 63\,\text{mm}$ calculated with Eq.\ref{eqn_21}. \ Second (continuous line), employing the same emitter as in Fig.\ref{eqn_5}, i.e., $R \simeq 31\,\text{mm}$ and $N_r=35$. All cases operate at $f_0=1\;\text{MHz}$ with $L=60\,\text{mm}$ and $N=6$.}
\label{fig_7}
\end{figure}

On the other hand, the $\theta_{\rm min}$ axicon angle and its corresponding depth $z_{\theta_{\rm min}}$, do not get changed; this is due to the way in which we assign the value of $Q$, by relation $Q={b\,\omega_0}/{c}-{2\pi N}/{L}$ with $b\to 1$ (e.g. b=0.998). \ In this way, the values of $Q$ are the minimum possible achieved with the current values of $N$ and $L$, ensuring that $|\boldsymbol{\Psi}(\rho, z=0)|$ does not oscillate excessively.

\begin{table}[h]
\caption{Parameters for the FW profiles shown in {Fig.\ref{fig_5}}(a).}
\hypertarget{tab4}{\label{tab_2}}
\small{\footnotesize{
\hspace{0.4mm}
\begin{tabularx}{9.0cm}{|c|c|c|c|c|c|}\hline
\begin{tabular}{c}Case \end{tabular} &
\begin{tabular}{c} $\theta_{\rm min}$\\ $[\text{deg}]$ \end{tabular} &
\begin{tabular}{c} $\theta_{\rm max}$\\ $[\text{deg}]$ \end{tabular} &
\begin{tabular}{c} $z_{\theta_{\rm min}}$\\ $[\text{mm}]$ \end{tabular} &
\begin{tabular}{c} $z_{\theta_{\rm max}}$\\ $[\text{mm}]$ \end{tabular} &
\begin{tabular}{c}\!\!\! ${\Delta}$\end{tabular}\\\hline
 $L=60\;\text{mm}$      & $2.6$ & $32.3$ & $489$ & $30$ & $0.36$ \\\hline
 $L=120\;\text{mm}$      & $2.6$ & $32.3$ & $693$ & $49$ & $0.17$ \\\hline
\end{tabularx}
}}
\end{table}

To conclude this Appendix, we like to compare our previous results, with attenuation, with those obtained when acoustic FWs were created in an ideal non-absorbing medium. To this aim, Fig.\ref{fig_7} illustrates the patterns obtained for the same FW  in
Figs.\ref{fig_5} [see Case 1 in Sec.7: Cf. Eq.(\ref{eqn_25})].

\hyphenation{diameter}
\noindent The difference, in this case, focuses on the size of the aperture used by the computer program for the creation of the FWs. The first pattern (dotted line) uses the conservative approach of Eq.(\ref{eqn_21}), i.e., with $R \simeq 63\,\text{mm}$ of
radius (that is, $\varnothing>12.5\,\text{cm}$ of diameter!). While the second (continuous line) employs the radius $R\simeq 31\,\text{mm}$ as in Figs.\ref{fig_5}.

In all cases (as well as in those of Figs.\ref{fig_5} and Fig.\ref{fig_6}), the
width $d=0.6\,\text{mm}$, and kerf $\Delta_d=0.3\,\text{mm}$ of the rings were the same.

Notice how, when using the minor radius $R\cong31\,\text{mm}$, the patterns, indicated by the black lines in figures \ref{fig_5}
 and \ref{fig_7}, are quite similar. That is, both are falling down before the theoretical FW (solid image) finishes. This happens even though the first includes the effect of attenuation, while the second does not. \ On the contrary, when using the bigger radius $R\simeq 63\,\text{mm}$ given by Eq.\ref{eqn_21}, the profile (dotted line) is enhanced, resembling the theoretical FW. \ It is also possible to observe near $z=65\,\text{mm}$ the beginning of the repetition of the original FW pattern  (solid image). This is of course the consequence of using a Fourier representation. This does not happen in {Fig.\ref{fig_5}}(b), because of the use of a bigger value of $L$, and also because of the effect of absorption. It can moreover be observed how the non-attenuated original FW pattern in figure \ref{fig_7}, with $5$ peaks, is (moderately) distorted to $3$ peaks as in Figs.\ref{fig_5}.

\

\

\

\

\

\end{document}